\documentclass[letterpaper,preprintnumbers,nofootinbib,noshowpacs,eqsecnum,prd,superscriptaddress]{revtex4}

\usepackage{graphicx}
\usepackage{amsmath,amssymb}
\usepackage{url}
\usepackage{ulem}
\usepackage{hyperref,color,wrapfig}
\usepackage{bm}

\makeatletter 

\renewcommand{\p@subsection}{}

\renewcommand{\p@subsubsection}{}

\makeatother

\graphicspath{{.}{figs_jpg/}{figs_pdf/}}

\newcommand{\ope}{\ensuremath{\mathcal{O}}}
\newcommand{\lag}{\ensuremath{\mathcal{L}}}

\newcommand{\cosw}[1][]{\ensuremath{\cos^{#1}\theta_W}}
\newcommand{\gesim}{\,{_{\textstyle >}\atop^{\textstyle\sim}}\,}

\begin{document}

\title{Precision Measurements of Higgs Couplings: \\
       Implications for New Physics Scales}

\author{C.~Englert}
\affiliation{SUPA, School of Physics and Astronomy, University of
  Glasgow, United Kingdom}
\author{A.~Freitas}
\affiliation{PITT-PACC, Department of Physics \& Astronomy,
    University of Pittsburgh, USA}
\author{M.M.~M\"uhlleitner}
\affiliation{Institut f\"ur Theoretische Physik, Karlsruhe Institute of Technology (KIT), Germany}
\author{T.~Plehn}
\affiliation{Institut f\"ur Theoretische Physik, Universit\"at Heidelberg, Germany}
\author{M.~Rauch}
\affiliation{Institut f\"ur Theoretische Physik, Karlsruhe Institute of Technology (KIT), Germany}
\author{M.~Spira}
\affiliation{Paul Scherrer Institut, Villigen, Switzerland}
\author{K.~Walz}
\affiliation{Institut f\"ur Theoretische Physik, Karlsruhe Institute of Technology (KIT), Germany}

\begin{abstract}

The measured properties of the recently discovered Higgs boson are in
good agreement with predictions from the Standard Model. However,
small deviations in the Higgs couplings may manifest themselves once
the currently large uncertainties will be improved as part of the LHC
program and at a future Higgs factory.	We review typical new physics
scenarios that lead to observable modifications of the Higgs
interactions.  They can be divided into two broad categories: mixing
effects as in portal models or extended Higgs sectors, and vertex loop
effects from new matter or gauge fields. In each model we relate
coupling deviations to their effective new physics scale. It turns out
that with percent level precision the Higgs couplings will be
sensitive to the multi-TeV regime.

\end{abstract}

\preprint{PSI--PR--14--01}
\preprint{KA--TP--06--2014}
\preprint{SFB/CPP--14--13}

\maketitle


\tableofcontents

\newpage

\section{Introduction}

The recent discovery of the Higgs boson~\cite{higgs} at the
LHC~\cite{ATLASCMS} can be considered as a triumph of quantum field
theory in describing the fundamental interactions between elementary
particles. The postulation of the Higgs boson defines the structure of
the electroweak Standard Model (SM) and is the key ingredient to its
renormalizability. While there might be good reasons to suspect that
there exist intermediate new physics scales to account for dark
matter, the quark flavor structure, neutrino masses, the baryon
asymmetry of the Universe, or a full gauge coupling
unification~\cite{bsm_reviews}, the Standard Model is structurally
complete. This means that we can, in principle, make statements about
Lagrangians describing physics at the GUT-scale using renormalization
group evolution from the weak-scale Lagrangian~\cite{rge_studies}. In
addition, the dominant production process
as well as the most significant decay mode in the
Higgs discovery are both induced by quantum effects. At the Born level
the Higgs couples neither to gluons nor to photons, and the existence
and the size of these loop-induced couplings is already a decisive
test of the (effective) Standard Model \cite{lecture}.\medskip

With current data, all properties of the observed new state turn out to be in
rough agreement with expectations of the Standard Model~\cite{ACprop}, but the
experimental uncertainties are still large. A refinement of this coarse picture
in the future may reveal deviations from the minimal SM scenario. The
determination of zero-spin and positive parity~\cite{choi1,angana}, required by
isotropy of the vacuum, and the determination of the Higgs couplings
to SM
particles~\cite{sfitter_higgs,Bechtle:2014ewa,cplggen,cplggen2},
gauge bosons and
leptons/quarks, are the agents probing the Higgs mechanism \textit{sui
  generis} for generating SM particle masses. The Higgs couplings are presently
constrained at the level of several tens of percent,
soon to be improved at the
LHC~\cite{sfitter_higgs,Bechtle:2014ewa,sfitter_ilc} to about
20\%\footnote{Note that at the LHC we can only measure ratios of couplings
without making model assumptions.}. The high-luminosity run of the LHC (HL-LHC)
will reduce the errors to about 10\%.  Measurements at a future $e^+e^-$ linear
collider (LC)~\cite{sfitter_ilc,Bechtle:2014ewa,ilc} can improve
the accuracy to about 1\%,
cf.~Table~\ref{tab:cplgs}. 
Combining HL-LHC and HL-LC results will not give a
significant improvement for most of the couplings. A notable exception is
$h\gamma\gamma$, which is statistics limited even at the high-luminosity LC
(HL-LC). The improved determination of the other couplings allows to better
exploit the potential of the diphoton final state at the LHC, and, by chance, leads
to the same precision for the $h\gamma\gamma$ and $hgg$ couplings.
\medskip

\begin{table}[b!]
\begin{tabular}{|l||c|c||c|c||c|}
\hline
coupling	    &  LHC  & HL-LHC	       & LC	   & HL-LC & HL-LHC + HL-LC \\
\hline\hline
$hWW$		    &  0.09 & 0.08	       & 0.011	   & 0.006 & 0.005	    \\
$hZZ$		    &  0.11 & 0.08	       & 0.008	   & 0.005 & 0.004	    \\
$htt$		    &  0.15 & 0.12	       & 0.040	   & 0.017 & 0.015	    \\
$hbb$		    &  0.20 & 0.16	       & 0.023	   & 0.012 & 0.011	    \\
$h\tau\tau$	    &  0.11 & 0.09	       & 0.033	   & 0.017 & 0.015	    \\
\hline
$h\gamma\gamma$     &  0.20 & 0.15	       & 0.083	   & 0.035 & 0.024	    \\
$hgg$		    &  0.30 & 0.08	       & 0.054	   & 0.028 & 0.024	    \\
\hline
$h_{\text{invis}}$   &	---  & ---		& 0.008     & 0.004 & 0.004	     \\
\hline
\end{tabular}
\caption{Expected accuracy at the 68\% C.L.\ with which fundamental and derived
  Higgs couplings can be measured; the deviations are defined as
  $g=g_{SM} [1\pm\Delta]$ compared to the Standard Model at the LHC/HL-LHC (luminosities
  300 and 3000 fb$^{-1}$), LC/HL-LC (energies 250+500~GeV / 250+500~GeV+1~TeV and luminosities
  250+500 fb$^{-1}$ / 1150+1600+2500 fb$^{-1}$), and in combined analyses of HL-LHC and HL-LC.
  For invisible Higgs decays we give the upper limit on the underlying
  couplings. Constraints on an invisible Higgs decay width
  involve model-specific assumptions at the LHC, see 
  {\sl e.g.}~\cite{Dobrescu:2012td}. Therefore, we allow for additional
  contributions to the total Higgs width only in the linear collider scenarios, where these
  can be constrained model-independently by exploiting the recoil measurement~\cite{ilc}.
  }
\label{tab:cplgs}
\end{table}

The interactions of the Higgs boson could deviate from their SM values
if the Higgs mixes with other scalars, if it is a composite particle
or a mixture between an elementary and composite state (partial
compositeness), or through loop contributions from other new
particles. Thus, precision measurements of Higgs properties are
sensitive to physics beyond the Standard Model, potentially residing
at scales much higher than the Higgs vacuum expectation value
(vev). Depending on the strength and type of coupling between the new
physics and the Higgs boson, limits derived from Higgs data may exceed
those from direct searches, electroweak precision measurements, or
flavor physics. This way Higgs precision analyses open a unique window
to new physics sectors that are not strongly constrained by existing
results. \medskip

One way to study physics beyond the Standard Model (BSM) in terms of a
well defined quantum field theory is given by the effective field
theory approach. By assuming a few basic principles, like the field
content and the gauge symmetries of the Standard Model, deviations
from the Standard Model are parametrized by higher-dimensional
operators. While this approach allows us to study a large class of
models it also has its limitations: for example, it cannot account for
effects that arise from light particles, whose contributions may be
enhanced in BSM models, or from Higgs decays into new non-SM
particles. Therefore, to give a complete picture of BSM effects in the
Higgs sectors we also study specific BSM models which capture such
features. \medskip

In this paper we review characteristic scenarios to describe modified
Higgs couplings from physics beyond the Standard Model, and illustrate
their phenomenology with a few representative concrete models. In
particular, we try to address the question what high-energy scales can
be probed by precise measurements of Higgs couplings. While the answer
to this question is necessarily model dependent, our aim is to work
with a few typical scenarios and models that are archetypal examples
for a much larger class of models. We start by introducing modified
Higgs couplings in an effective theory philosophy in
Section~\ref{sec:eff}.  After that, we analyze the relationship
between Higgs coupling deviations and the scale of new physics in two
broad categories: modified Higgs properties through mixing effects
(Section~\ref{sec:mix}) and through loop effects
(Section~\ref{sec:loopeffects}). In Section~\ref{sec:sum}, we evaluate
and summarize the sensitivities to high scales in the different new
physics scenarios.

\section{Effective interactions}
\label{sec:eff}

In the present article we will give a survey of typical scales of new physics
beyond the Standard Model which can be probed in precision
measurements of the Higgs couplings. Deviations from the SM values are
predicted in many scenarios, of which a few representative examples
will be described later in detail. Unless the underlying model
violates the decoupling theorem \cite{appcar}, operator
expansions~\cite{BuWy,53op} suggest deviations of the order of
\begin{equation}
   g = g_\text{SM} \; [1 + \Delta] \;\; : \;\; \Delta = \ope(v^2 / \Lambda^2) \; ,
\label{eq:def_coups}
\end{equation}
with $v \approx 246$~GeV denoting the vacuum expectation value of the
standard Higgs field and $\Lambda \gg v$ the characteristic scale of
physics beyond the Standard Model.
For typical examples of models which violate the decoupling theorem see
Refs.~\cite{dcpl2,g4} and our discussion in Section~\ref{sec:stb}.

A theory with Standard Model operators but free Higgs couplings according to
Eq.\eqref{eq:def_coups} is neither unitary nor electroweak
renormalizable~\cite{eft}.
However, it can be regarded as an effective theory which contains additional
higher-dimensional operators suppressed by powers of $\Lambda$. The
effective model can be thought of as emerging from a complete and
UV-consistent fundamental theory with a decoupled heavy sector. One
such completion could be a Two-Higgs-Doublet model (2HDM)
\cite{2hdm,dcpl} with all
heavy Higgs masses around the high scale $\Lambda$, well
separated from the light Higgs mass (see end of this section).

According to Eq.\eqref{eq:def_coups}
experimental accuracies of $\Delta = 0.2$ down to 0.01 will give us sensitivity to scales
of order $\Lambda \sim$ 550~GeV up to 2.5~TeV.
While the smaller of the two bounds is complementary
to direct LHC searches, the larger of the two bounds
generally exceeds the direct search range of LHC. Thus precision
measurements in the Higgs sector may allow us to enter
new physics territory.\medskip

A system in which the particles obey the symmetry structure of the Standard Model,
but supplemented by interactions which are generated at high scales
$\Lambda \gg v$, can be described by the effective interaction
\begin{equation}
\lag_\text{eff} = \sum^{6,...}_{D=2} \dfrac{1}{\Lambda^{D-4}} \; \lag_D \;.
\label{eq:def_eft}
\end{equation}
Here $D$ characterizes the mass dimensions of the various terms of the
effective Lagrangian. Higgs mass terms carry dimension $D =
2$ and minimal interaction terms $D
= 4$. In this picture, large Higgs masses of ${\cal
O}(\Lambda)$ can only be avoided either by fine-tuning operators of dimension $D=2$
or by introducing new symmetries, as for example in supersymmetric
extensions or Little Higgs scenarios.
For the analysis of any higher-dimensional system of the kind
sketched in Eq.\eqref{eq:def_eft} it is crucial that one defines a
complete operator basis and keeps in mind which set of operators a given
coupling measurement corresponds to. As an example, the results of the
Higgs couplings fit based on the $D \le 4$ Lagrangian will change when
we include a free Higgs coupling to photons or gluons at $D=6$. The
measured central values and error bars, for example of the top Yukawa
coupling, are critically affected by this change, so that every Higgs
coupling extraction is defined in relation to a unique set of operators
in the Lagrangian.\medskip

The effects of high-scale physics on the SM Higgs field and its interactions can
be categorized in two classes: (i) mixing effects of the Higgs field
with other high-mass scalar fields, and (ii) vertex effects modifying
the couplings between the Higgs field and gauge bosons, quarks and leptons, and
Higgs self-interactions.

\begin{enumerate}
\item \underline{Mixing effects:} The standard Higgs field may mix
  with other scalar fields. The operators describing the mixing
  effects carry mass dimension $D = 4$ in the combined SM/new scalar
  system, reducing to $D = 2$ in the effective SM Lagrangian after
  symmetry breaking. There, mixing reduces the mass of the Higgs boson by
  $\ope(\eta^2)$, with $\eta$ denoting the coupling between the SM
  Higgs boson and the new scalars. The mixing also modifies the
  strength with which the field couples to SM particles.  In basic
  portal models, in which the SM Higgs field is coupled with a hidden
  sector~\cite{Bij,Wells1,Bock,Wells2}, the couplings are
  reduced universally. The decays of SM particles into states of the
  hidden sector demand proper control of invisible Higgs decays if
  this scenario should be described
  conclusively~\cite{Englert:2011us}. A generic weakly interacting
  extension of the simplest Higgs sector includes a second Higgs
  doublet, as required in many models for physics beyond the Standard
  Model.  Alternatively, the Higgs sector can be strongly interacting
  and connected to theories of extra dimensions by an AdS/CFT
  correspondence~\cite{contino,Agashe:2004rs,Contino:2006qr,GroMMM,Bock}.
  Finally, analyses in the decoupling regime~\cite{dcpl} for large
  CP-odd pseudoscalar Higgs masses may open windows to areas not
  accessible in direct searches. It turns out that in this class energy
  scales with BSM physics can eventually be probed at the multi-TeV level.

\item \underline{Loop effects:} Vertex corrections of Higgs couplings
  to SM particles can be generated by virtual contributions of new
  gauge bosons, scalars or fermions, colored or non-colored. Typical
  examples are predicted in a large variety of models, for example
  supersymmetry, strong dynamics, extra dimensions, see-saw models, or
  extended gauge groups.  The vertex effects which are generated by
  the exchange of new heavy fields carry dimensions $D \ge 6$ for an
  $SU(2)$ doublet Higgs field.	Such loop effects come with
  suppression factors $1/(16\pi^2)$ in addition to potentially small
  couplings between the Standard Model and the new fields. Thus, only
  new mass scales not in excess of about $M < v/(4\pi \sqrt{\Delta})
  \sim$ 200 GeV can be probed, in most models much less than the
  direct search reach at the LHC.  As a result, loop effects are less
  promising for exploring new physics scales indirectly.
\end{enumerate}
\medskip

\begin{figure}
\raisebox{-2mm}{\includegraphics[width=0.42\textwidth]{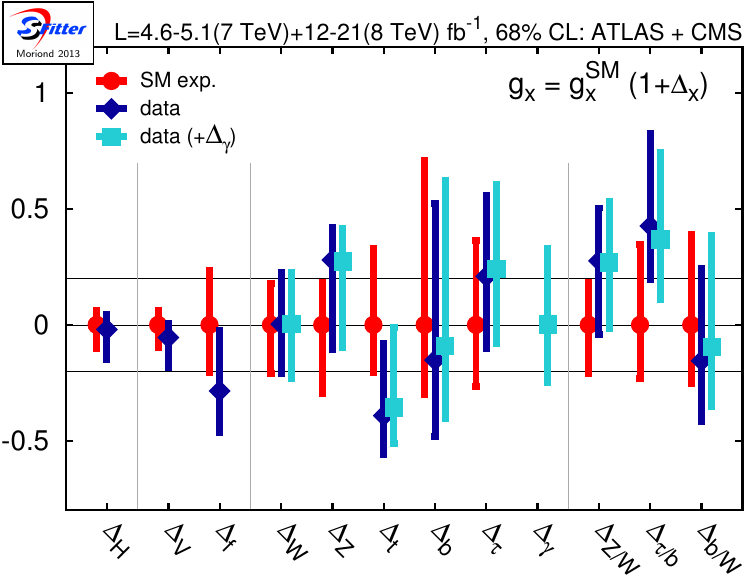}}
\hspace{0.1\textwidth}
\includegraphics[width=0.42\textwidth]{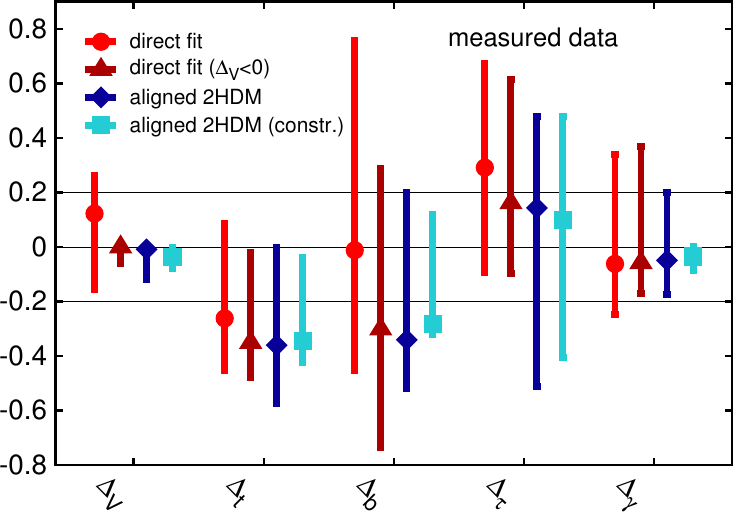}
\caption{Higgs coupling measurement based on all currently available
  ATLAS and CMS data. In the left panel we compare the SM expectation with a fit
  to the weak-scale Higgs Lagrangian with free couplings to the data, and either
  including a Higgs-photon coupling or not. In the last three
  columns we show errors on ratios of couplings, where, analogous to
  Eq.\eqref{eq:def_coups}, $\Delta$ parametrizes the deviation from the
  corresponding SM ratio.
  In the right panel we compare the fits to the
  weak-scale couplings with a fit to the aligned 2HDM in terms of
  the light Higgs couplings. Figures from Ref.~\cite{dcpl2}.
  The only difference between the cyan results in the left panel and the
lighter red ones in the right panel is that for the latter we set $\Delta_W =
\Delta_Z \equiv \Delta_V$.
}
\label{fig:sfitter}
\end{figure}

Before we discuss the effects of specific modifications of the
SM Higgs sector, we briefly review the approach taken in
most of the recent Higgs coupling analyses~\cite{dieter,
ATLASCMScoup,cplggen,cplggen2,sfitter_higgs,Bechtle:2014ewa}. In
these approaches, one introduces
a number of free couplings in the SM Lagrangian, corresponding to the number of independently measured
production and decay channels. In
the left panel of Fig.~\ref{fig:sfitter} we show such a coupling
fit for the maximum number of currently accessible couplings
as well as reduced sets, using the program SFitter~\cite{sfitter_higgs,dcpl2}. This provides a non-trivial
test of the Standard Model, in which all Higgs couplings are predicted by the
minimal realization of the Higgs mechanism.

At the LHC, Higgs decays into gauge bosons are very well established
through their characteristic final--state signatures~\cite{ATLASCMS}
and are the most precisely measured Higgs channels.  Higgs decays to
tau leptons can be identified in associated Higgs production with a
hard jet or in weak boson
fusion~\cite{Ellis:1987xu,signal_tau,ATLASCMS_tau},
while decays to bottom quarks can be measured in associated $WH$ and $ZH$
production~\cite{signal_b,ATLASCMS_vh}. From these channels, one can obtain
information about the Higgs couplings to photons, $W/Z$-bosons, tau
leptons and bottom quarks. However, because current $H \to b\bar{b}$
measurements are not sufficiently sensitive to probe Standard Model
coupling strengths, the information on the bottom Yukawa coupling in
the SFitter analysis~\cite{sfitter_higgs} is dominated by the
observed total Higgs production rate. Finally, the information on the
top Yukawa coupling comes entirely from the effective Higgs--gluon and
Higgs--photon couplings. This is why in Fig.~\ref{fig:sfitter} we can
choose to either show $\Delta_t$ or $\Delta_g$ as measured couplings.
In the future, we should be able to extract $t\bar{t}H$ production,
for example with a Higgs decay into bottom jets or $W$
bosons~\cite{signal_t}. Recently, some studies have suggested to
extract the top Yukawa coupling from the effective Higgs--gluon
coupling by resolving the top loop in the boosted Higgs
regime~\cite{resolveloop}.\medskip

When interpreting these Higgs coupling measurements we need to keep in mind that an arbitrary
modification of these couplings violates the
ultraviolet properties of the Standard Model, like renormalizability
and unitarity. This problem can be solved if we consider
the SM Lagrangian with free Higgs couplings as an effective
theory, which at some energy scale is completed by a weakly
interacting renormalizable field theory. The only condition on this
ultraviolet completion is that its free parameters allow a
free variation of all Higgs couplings in the Standard Model.  As an
example, we can interpret the current Higgs coupling measurements in
terms of an aligned Two-Higgs-Doublet model, where the Yukawa couplings of the
two Higgs doublets are proportional to each other in flavor space.
For simplicity, we
assume custodial symmetry ($\Delta_Z = \Delta_W \equiv \Delta_V < 0$),
which can be broken through loop effects at the percent level~\cite{ewfit}. At
tree level the aligned 2HDM has five free parameters, including the
mass of the charged Higgs boson contributing to the effective
Higgs--photon coupling. To eventually allow for a free Higgs--gluon
coupling it would have to be supplemented for example by a top partner
state.	In the right panel of Fig.~\ref{fig:sfitter} we compare the
extracted free Higgs couplings with the corresponding fit to the
aligned 2HDM parameters, translated into the SM coupling
deviations. We see that the central values as well as the error bars
agree well between these two approaches. Slight deviations arise because the complete model can
induce correlations between the couplings. If the aligned 2HDM is the true
underlying model, additional constraints arise from non-standard Higgs
searches and electroweak precision and flavor constraints.
These have been ignored for the blue error bands, but are additionally
taken into account for the cyan ones.

\subsection{Dimension-6 Lagrangian}
\label{sec:dim6}

Assuming the new physics sector to be $SU(3) \times SU(2) \times U(1)$
gauge invariant, the leading effect of a heavy new physics sector on the SM
Higgs field is described by effective $D = 6$
operators~\cite{BuWy,53op,Winter,Ebo,other}. For the introduction of scalar
singlet states $D = 5$ operators arise beyond the corresponding renormalizable
dimension-4 Lagrangian. Here we only consider an effective theory including
the Higgs isodoublet $\phi$ supplemented by all other SM particles.
The mass dimension of $\phi$
automatically induces a power counting in terms of a new energy
scale $\Lambda$~\cite{Brivio:2013pma} so that the effective Lagrangian
can be written as
\begin{equation}
{\cal L}_{\text{eff}} = \sum_n \frac{f_n}{\Lambda^2} \; {\cal O}_n \;,
\end{equation}
with the couplings denoted by $f_n$. A list of the operators ${\cal
  O}_n$ relevant to Higgs phenomenology can be found in Table~\ref{tab:D6op}.
The complete, but not minimal
list includes three types of operators: pure Higgs operators,
Higgs--gauge boson operators, and Higgs--fermion operators. These operators,
written in terms of the $SU(2)_L$ doublet $\phi$, are defined in the
linear representation of the Higgs and Goldstone fields. In this form
the Lagrangian shows the full electroweak symmetry structure before
electroweak symmetry breaking.	The implicit physics assumption behind
using this representation is that the particle discovered at the LHC is a
Standard-Model-like Higgs boson, where deviations from the Standard
Model case can be induced by mixing effects or an effective field
theory description of an unknown ultraviolet completion.\medskip

\begin{table}
\begin{tabular}{|l|l|l|}
\hline
Higgs--gluon
  & $\ope_{GG} = \phi^\dagger \phi\, \text{tr}\{G^2\}$
  & SM Higgs phenomenology  \\ \hline\hline
Higgs--vector boson (1)
  & $\ope_{\phi 1} = (D \phi)^\dagger \phi \phi^\dagger (D \phi)$
  & custodial symmetry violation  \\
  & $\ope_{\phi 4} = (D \phi)^\dagger (D \phi) \phi^\dagger \phi$
  & \\ \hline
Higgs--vector boson (2)
  & $\ope_{WW} =  \phi^\dagger W^2 \phi,\;\; \ope_{BB} $
  & SM Higgs decays $h \to \gamma \gamma, \gamma Z$  \\
  & $\ope_{BW} =  \phi^\dagger BW \phi$ & custodial symmetry violation \\
  & $\ope_W = (D \phi)^\dagger W (D \phi), \;\; \ope_{B}$
  & \\ \hline\hline
Higgs--fermion (1)
  & $\ope_{LR} =  (\phi^\dagger \phi) ({\bar{L}} \phi R)$
  & corrections to Yukawa couplings \\
  & $ \ope_{LL1} =  \phi^\dagger (i \overleftrightarrow{D} \phi)
  ({\bar{L}} \gamma L), \;\; \ope_{RR1}$
  & neutral current contributions \\
  & $\ope_{LL3} =  \phi^\dagger (i\overleftrightarrow{D}^a \phi)
  ({\bar{L}} \gamma \tau^a L)$
  & neutral/charged current contributions  \\ \hline
Higgs--fermion (2)
  & $\ope_{\phi B} =  \phi \bar{L} (\sigma B) R, \;\; \ope_{\phi W}, \;\; \ope_{\phi G}$
   & electric/magnetic moments \\ \hline\hline
Higgs self-coupling
  & $\ope_{\phi 2} = \tfrac{1}{2} | \partial (\phi^\dagger \phi) |^2$
   & weak boson fusion, decays $h\to VV$ \\
  & $\ope_{\phi 3} = \tfrac{1}{3} | \phi^\dagger \phi |^3 $
   & Higgs self-interactions \\
\hline
\end{tabular}
\caption{$D = 6$ operators of an $SU(3) \times SU(2) \times U(1)$
invariant theory beyond the Standard Model involving the SM Higgs field.
Notation: $\phi$ is the $SU(2)_L$ doublet; $W =
W_{\mu\nu}=i\frac{g}{2}\sigma^aW^a_{\mu\nu}$, $B =
B_{\mu\nu}=i\frac{g'}{2} B_{\mu\nu}$, $G = G_{\mu\nu}=i\frac{g_s}{2}
\lambda^a G^a_{\mu\nu}$ field strengths; $\partial$ space-time
derivative, $D$ covariant derivative; $\phi^\dagger
\overset\leftrightarrow{D} \phi = \phi^\dagger D \phi - (D \phi)^\dagger
\phi$; $L$ lepton/quark isodoublet, $R$ lepton/quark isosinglet; 1 =
isoscalar coupling, 3 = isovector coupling; and $\gamma=\gamma_\mu$,
$\sigma = \sigma_{\mu\nu}$, $\sigma^a$ Pauli matrices, $\lambda^a$
Gell-Mann matrices. Scale parameters $f_{GG}/\Lambda^2$ {\sl etc.}\ define
the impact of the operators. The conventions of Ref.~\cite{Ebo} can be
obtained by the identifications $\ope_{LR}\to \ope_{f\phi}, $
$\ope_{LL1,RR1} \to \ope^{(1)}_{\phi f}$, $\ope_{LL3} \to
\ope^{(3)}_{\phi f}$.}
\label{tab:D6op}
\end{table}

Table~\ref{tab:D6op} contains a rich chirality, isospin and flavor
structure in higher-dimensional Higgs--fermion operators, scaled by
$f_{GG}/\Lambda^2$ {\sl etc.} This includes the usual scalar and vector
currents, but also the dipole operators like $\ope_{\phi B}$ {\sl etc.}
The operators in Table~\ref{tab:D6op} are not independent but related
by the [classical] equations of motion. Three equations connect the
Higgs, gauge and fermion operators in a non-trivial
manner~\cite{53op,Ebo}, in standard notation for the couplings and
  with the hypercharges for left-handed and right-handed fermions denoted by
  $Y_L$ and $Y_R$:
\begin{alignat}{5}
\ope_{\phi 4}
   &= - \ope_{\phi 2} + \frac{1}{2}\,\sum_{\ell , q} (y \,
     \ope_{LR} + \text{h.c.}) - \frac{1}{2}\, \frac{\partial V(h)}{\partial h} \notag \\
\ope_B &= - \frac{1}{2}\,\ope_{BW} - \frac{1}{2}\,\ope_{BB}
   - \frac{{g'}^2}{2}\, \ope_{\phi 1} + \frac{{g'}^2}{4} \ope_{\phi 2}
   -   \frac{{g'}^2}{4} \sum_{\ell, q}
     \left( Y_L \ope_{LL1} + Y_R \ope_{RR} \right)	  \notag \\
\ope_W &= - \frac{1}{2}\,\ope_{BW} - \frac{1}{2}\,\ope_{WW}
     - \frac{g^2}{2}\ope_{\phi 4} + \frac{g^2}{4}\ope_{\phi 2}
     - \frac{g^2}{8} \sum_{\ell, q} \ope_{LL3}	\; .
\label{eq:eom}
\end{alignat}
The summed indices include the full generation structure of the
Higgs--fermion operators. We are free to use these equations to
eliminate three higher-dimensional Higgs operators of our choice.
The discussion of the operators relevant for the purpose of this review
becomes most transparent when we use these relations to directly
eliminate $\ope_{\phi 4}, \ope_{B}$ and $\ope_{W}$. \medskip

All operators given in Table~\ref{tab:D6op} respect the gauge symmetry
structure of the Standard Model. However, the Standard Model has additional
(accidental) global symmetries motivated by experimental observations and
with phenomenological implications.
In the Higgs sector, custodial symmetry is broken at the loop level by
the mass splitting of the fermion isodoublets and by gauging
hypercharge as a subgroup of the bigger $SU(2)_R$ global symmetry of
the SM Higgs sector, which guarantees $m_W/m_Z=\cosw$ for bare
quantities.
However, some of the $D=6$ operators lead to additional custodial breaking
contributions, which can be described by the $S$ and $T$ parameters~\cite{pt}.
At tree level one obtains~\cite{st},
\begin{alignat}{5}
  \alpha \Delta S &= -e^2 v^2 \; \frac{f_{BW}}{\Lambda^2} \,,
  \qquad\quad
  \alpha \Delta T &= -\frac{v^2}{2} \; \frac{f_{\phi 1}}{\Lambda^2} \, .
\end{alignat}
Moreover, $\ope_{BB}, \ope_{WW}, \ope_{B}, \ope_{W}$ generate
tree-level contributions to the extended set of oblique parameters
$Y,W$~\cite{Barbieri:2004qk}. Compared
to one-loop contributions in the Standard Model, which have been very
successful in predicting the top and Higgs masses~\cite{ewfit} these
contributions are not necessarily small. These six operators, together with
$\ope_{LL1}$, $\ope_{LL3}$ and $\ope_{RR1}$, are strongly constrained by
$Z$-pole measurements and bounds on anomalous gauge boson interactions from
$W^+W^-$ production at LEP2 \cite{cplggen2,Barbieri:2004qk,tripl,pomarolriva,Contino:2013kra}.
However, there are not enough independent electroweak precision observables to
obtain separate bounds on all operators in this list. Therefore we here neglect
$\ope_{\phi 1}$, $\ope_{BW}$, $\ope_{LL1}$, $\ope_{LL3}$ and $\ope_{RR1}$, but
keep $\ope_{BB}$ and $\ope_{WW}$, which allow a significant deviation of the
decay $h \to Z
\gamma$ from the SM prediction \cite{zgamma}. This simple choice is adequate for
current Higgs measurement uncertainties. In principle, however, one has to
consider all dimension-6 operators contributing to 
electroweak precision data at tree-level and carefully map out
cancellations between them. Such
cancellations are known to happen for example for vector resonances,
kinetic mixing, or additional fermionic
matter~\cite{g4,Lavoura:1992np,stfermioncontr,sew,Gillioz:2013pba}. \medskip

The dipole operators $\ope_{\phi B}$ {\sl etc.}\ are strongly
constrained by measurements of the electric dipole moments and the
anomalous magnetic moment, so that their contributions to the decay rates of the
Higgs boson into fermion pairs can be neglected compared to $\ope_{LR}$.
For the third generation such dipole operators
can eventually be tested in $t\bar{t}h$ and $b\bar{b}h$
production~\cite{cdm}, which are production channels we do not
consider here. Exploiting the equations of motion in Eq.\eqref{eq:eom},
we are now left with the reduced operator basis
\begin{equation}
\{ \; \ope_{GG}, \; \ope_{WW}, \; \ope_{BB};
   \; \ope_{\phi 2}, \; \ope_{\phi 3};
   \; \ope_{LR} \;
\}
\label{eq:d6_basis}
\end{equation}
for the dimension-6 Higgs operators analysis.  \medskip

The set of operators given in Eq.\eqref{eq:d6_basis} is a basis of
the leading higher-dimensional operators given the symmetry structure
of the ultraviolet completion of the Standard Model and our choice of
using the equations of motion. However, there are many ways to reduce
this dimension-6 operator basis by making additional assumptions about
the (experimentally unknown) model and its symmetry structure. From a
phenomenological perspective such assumptions are not helpful.
Instead, data from LHC and from a future $e^+e^-$ linear collider should
help us determine the structure of the Higgs sector based on the most
general possible analysis.  The only justification for additional
simplifications can be fundamental shortcomings of the available data, for
example a common lack of distinguishing power of collider searches.
While we do not see how such an argument can be used in the gauge
sector, we will resort to it for the Higgs couplings to fermions.
\medskip

The dominant effect of higher-dimensional Higgs operators are modified
relations between the dimension-4 Higgs potential ($\mu,\lambda$) and
the main observables $m_h$ and $v$. In addition, shifts in the Higgs
wave function renormalization in general affect triple and quartic
Higgs couplings, as well as an additional universal modification of
the gauge boson--Higgs and fermion--Higgs couplings.  The derivatives
or momentum-dependent self interactions induced by $\ope_{\phi 2}$
indicate strong self-interactions in the regime where the energy of
the scattering process gets close to the suppression scale $\Lambda$.
The price we pay for applying an effective
  field theory approach is that multiple Higgs couplings are
generated with the same suppression factor $f_{\phi
  2}/\Lambda^2$~\cite{silh}. Strong coupling effects for example from
$\ope_{\phi 2}$ can be observed experimentally as a significant rate
enhancement in Higgs pair production
\cite{silh,Grober:2010yv,Contino:2010mh,Contino:2013gna,Dolan:2012ac}
compared to single Higgs production at the LHC. In contrast,
$\ope_{\phi 3}$ will merely affect the value of the triple and quartic
Higgs couplings at the LHC.

\subsubsection*{Higgs self-interactions}

Since our main focus is not on the description of multi-Higgs
interactions, we will adapt a canonical normalization of the Higgs
kinetic term in the following. This implies a universal shift~\cite{BargerZer,Ebo}
\begin{equation}
  \label{eq:shift}
  h\to \left[1 - \frac{v^2}{4\Lambda^2}
  (f_{\phi 1} + 2f_{\phi 2} + f_{\phi 4})
  \right]h + \mathcal{O} ({v^4/\Lambda^4})\,,
\end{equation}
and the renormalization factors have to be included in the Higgs
couplings. The link between the operators of Table~\ref{tab:D6op} and
observable Higgs interactions is given by the corresponding effective
Lagrangian. Note that the induced shift in the Higgs mass term merely
re-defines the bare Higgs mass, rather than inducing an observable
effect.
\medskip

We can write the Lagrangian containing solely pure Higgs
interactions~\cite{BargerZer} to track the effects of the Higgs
operators in Table~\ref{tab:D6op}.
\begin{alignat}{5}
\lag_\text{eff}^{h} =
&-
 \frac{m_h^2}{2v}\left[
 \left(1-\frac{f_{\phi 124} \; v^2}{2\Lambda^2}
 - \frac{2 f_{\phi 3}\; v^4}{3\Lambda^2 m_h^2} \right) h^3
 -\frac{2f_{\phi 124} \; v^2}{\Lambda^2 m_h^2}
 h \, \partial_\mu h \, \partial^\mu h\right]
 \notag \\
&-
 \frac{m_h^2}{8v^2}\left[\left(1-\frac{f_{\phi 124} \; v^2}{\Lambda^2}
 -\frac{4f_{\phi 3} \; v^4}{\Lambda^2 m_h^2}\right) h^4
 -\frac{4f_{\phi 124} \; v^2}{\Lambda^2 m_h^2} h^2 \, \partial_\mu  \,
 h\partial^\mu h\right] \,,
\label{eq:lag6_higgs}
\end{alignat}
where $f_{\phi 124} \equiv \frac{1}{2}f_{\phi 1} + f_{\phi 2} + \frac{1}{2}f_{\phi 4}$.
For completeness, we have included the contribution of $\ope_{\phi
1}$, but will disregard it in the following, since it is strongly constrained by
electroweak precision data. Also, at this point we have still retained the 
operator $\ope_{\phi 4}$, which can be eliminated through the equations of
motion in Eq.\eqref{eq:eom}.
The corrections to the Higgs self-couplings appear in two distinct
patterns. First, the Standard Model coupling strengths are modified by
corrections of the form $v^2/\Lambda^2$ or $v^4/(m_h^2 \Lambda^2)$, both
of which are suppressed as long as $\Lambda \gg v$. Second, after Fourier
transformation the last term in each line gives rise to modifications
proportional to $p^2/\Lambda^2$. They are only small as long as the
given observable probes small momentum scales $p \ll \Lambda$.
Observables which probe a range of energies, such as
longitudinal gauge boson scattering, will be
dominated by physics at larger scales. If these are close to
the cut-off scale we need to include an appropriate matching condition
to the ultraviolet completion. However, at the LHC
contributions from larger scales are usually suppressed by the parton
densities, so that many observables are not sensitive to the particular
structure of the ultraviolet completion.
\medskip

\subsubsection*{Higgs--gauge boson interactions}

The same translation of the operators listed in Table~\ref{tab:D6op} to
an effective Lagrangian for the Higgs--gauge sector reads
\begin{alignat}{7}
\lag_\text{eff}^{hVV}
&= g_{hgg} \; h G^a_{\mu\nu} G^{a\mu\nu}
 + g_{h\gamma\gamma} \; h A_{\mu \nu} A^{\mu \nu} \notag \\
&+ g^{(1)}_{hZ\gamma} \; A_{\mu \nu} Z^\mu \partial^\nu h
 + g^{(2)}_{hZ\gamma} \; h A_{\mu \nu} Z^{\mu \nu} \notag \\
&+ g^{(1)}_{hZZ} \; Z_{\mu \nu} Z^\mu \partial^\nu h
 + g^{(2)}_{hZZ} \; h Z_{\mu \nu} Z^{\mu \nu}
 + g^{(m)}_{hZZ} \; h Z_\mu Z^\mu \notag \\
&+ g^{(1)}_{hWW} \; \left (W^+_{\mu \nu} W^{- \mu} \partial^{\nu} h +\text{h.c.} \right)
 + g^{(2)}_{hWW} \; h W^+_{\mu \nu} W^{- \, \mu \nu}
 + g^{(m)}_{hWW} \; h W^+_\mu W^{- \mu} \; ,
\label{eq:lag6_gauge}
\end{alignat}
with $V_{\mu\nu}=D_\mu V_\nu - D_\nu V_\mu$ ($V=B,W,G$)
supplemented by the corresponding rotations to the mass eigenstates
$W^\pm,Z,A$.
The coupling strengths of the symmetric,
higher-dimensional operators can be related to the Wilson coefficients
of the effective Lagrangian of the broken theory as~\cite{Ebo,Hagiwara:1993qt}
\begin{alignat}{7}
g_{hgg} &= -\frac{g_s^2 v}{2\Lambda^2} \; f_{GG}
 &\quad
g_{h\gamma\gamma} &= - \frac{g^2 v s_W^2}{ 2\Lambda^2} \; \frac{f_{BB} + f_{WW} - f_{BW}}{2}
 \notag \\
g^{(m)}_{hZZ} &= \frac{g m_Z}{2 c_W} \,
		 \left[1+\frac{v^2}{2\Lambda^2}\left( f_{\phi 4}-f_{\phi 2}
		 +\frac{f_{\phi 1}}{2} \right) \right]
 &\quad
g^{(m)}_{hWW} &= g m_W \,
		 \left[1+\frac{v^2}{2\Lambda^2}\left( f_{\phi 4}-f_{\phi 2}
		 -\frac{f_{\phi 1}}{2} \right)\right]
 \notag \\
g^{(1)}_{hZZ} &= \frac{g^2 v}{ 2\Lambda^2} \; \frac{c_W^2 f_W + s_W^2
f_B}{2 c_W^2}
 &\quad
g^{(2)}_{hZZ} &= - \frac{g^2 v}{2\Lambda^2} \;
		   \frac{s_W^4 f_{BB} +c_W^4 f_{WW} + c_W^2 s_W^2
f_{BW}}{2 c_W^2}
 \notag \\
g^{(1)}_{hWW} &= \frac{g^2 v}{2\Lambda^2} \; \frac{f_W}{2}
 &\quad
g^{(2)}_{hWW} &= - \frac{g^2 v}{2\Lambda^2} \; f_{WW}
 \notag \\
g^{(1)}_{hZ\gamma} &= \frac{g^2 v}{2 \Lambda^2} \; \frac{s_W (f_W - f_B)
}{2 c_W}
 &\quad
g^{(2)}_{hZ\gamma} &= \frac{g^2 v}{2 \Lambda^2} \;
		     \frac{s_W [2 s_W^2 f_{BB} - 2 c_W^2 f_{WW} +
(c_W^2-s_W^2)f_{BW} ]}{2 c_W}
 \; ,
\label{eq:lag6_gauge2}
\end{alignat}
where $s_W$ and $c_W$ denote the sine and cosine of the weak mixing
angle.
As in Eq.~\eqref{eq:lag6_higgs}, we also show how the contributions of $\ope_{\phi 1}$
and $\ope_{BW}$ enter in the formulae, but will continue to ignore these
operators in the following.
The effective scale parameter $\Lambda$ includes all possible couplings and
loop factors of the kind $1/(16\pi^2)$.
In Section~\ref{sec:examples} we will see that in specific models, where
these additional factors are known, the bounds on the actual mass
scale can be significantly weaker than in the general form of
Eq.\eqref{eq:lag6_gauge} due to loop suppression factors.\medskip

In these general expressions, Eqs.\eqref{eq:lag6_higgs} and
\eqref{eq:lag6_gauge2}, we have not yet made use of the
equations of motion in Eq.\eqref{eq:eom}, because there is no general
agreement which three operators to remove with their help. The
couplings $g^{(m)}_{hVV}$ are linked to the heavy gauge boson masses
and already exist in the renormalizable dimension-4 SM Lagrangian.
To arrive at our basis of Eq.\eqref{eq:d6_basis} we have to replace
the coupling factors $f_j$ by a new set $f'_j$, where
\begin{alignat}{5}
f'_{\phi 4} = f'_B = f'_W = 0 \; .
\end{alignat}
The remaining, finite coupling factors are shifted accordingly \textsl{e.g.}
$f'_{BB} = f_{BB} - f_B/2 ...$, {\sl etc.}  The couplings
$g_{hgg}$ and $g_{h\gamma\gamma}$ are generated by dimension-6
operators, but play a special role both in experiment and in
theory. First, in spite of being quantum effects they are the basis of
the Higgs discovery at the LHC. Second, they are generated by SM
states running in a closed loop and the corresponding
operators are not suppressed by a heavy mass scale like $\Lambda =
m_t$, see also Section~\ref{sec:loopi}.
The Yukawa coupling between the Higgs boson and a massive chiral
fermion in $f_{GG}$ circumvents the Appelquist--Carazzone decoupling
theorem~\cite{appcar}. It leads to a scaling $f_{GG} \sim \Lambda^2/v^2$ which
cancels the explicit $1/\Lambda^2$ suppression and replaces it by
$1/v^2$. This non-decoupling behavior is the basis for the
experimental exclusion of a chiral fourth
generation~\cite{g4,g4a}.\medskip

The alert reader might realize that modifying the $hWW$ and $hZZ$
couplings as indicated in Eq.\eqref{eq:lag6_gauge2} can lead to
unitarity violation within the effective theory picture already for
scales $p^2<\Lambda^2$. This modification in longitudinal gauge boson
scattering will be compensated by new heavy scalar or vector resonances,
which are integrated out and  thus cannot be accounted for
in the effective theory. Searches for such states have been
described for example in Ref.~\cite{fill} and can be considered independent
of the Higgs measurements as long as the narrow width
approximation is valid.\medskip

The labeling of the couplings in Eq.\eqref{eq:lag6_gauge2} correctly
suggests that there are many ways to modify a Higgs coupling like
$g_{hWW}$ through higher-dimensional operators. Such coupling shifts
can arise from a non-standard renormalizable coupling as well as
different dimension-6 operators. Based on rate measurements these
effects cannot be distinguished. One way to gain some insight into the
source of a possible deviation of LHC or linear collider measurements
from the Standard Model prediction are additional constraints on the
same set of dimension-6 operators, for example from anomalous gauge
couplings or electroweak precision
data~\cite{Ebo,Hagiwara:1993qt,Alloul:2013naa}. Another way to
separate different anomalous couplings are distributions, preferably
in Higgs production processes which involve more particles than the Higgs
boson. In that case the energy dependence or the Lorentz structure of
an operator will be reflected in angular correlations or transverse
momentum spectra of different particles
produced~\cite{Ellis:1987xu,Baur:1989cm,Brivio:2013pma,Contino:2013kra,
Alloul:2013naa,dist}.

\subsubsection*{Higgs--fermion interactions}

In our illustrative analysis we are mainly concerned with collider
measurements, where these couplings mediate flavor-diagonal fermionic
decays assuming some kind of minimal flavor violation. For a
review of flavor-violating objects we refer to
Ref.~\cite{Isidori:2010kg}. Moreover, we will only consider
operators where current constraints from non-Higgs observables
leave room for appreciable modifications of the Higgs branching ratios.
Therefore, we limit ourselves to
modifications of the fermion masses or Yukawa couplings,
\begin{alignat}{7}
\lag_\text{eff}^{hff} \simeq
&-\frac{h}{\sqrt{2}} \sum_{b,t,\tau} \bar{L}
  \left( y - \frac{v^2}{2 \Lambda^2} f'_{LR} \right) R
 +\text{h.c.}
\label{eq:lag6_fermion}
\end{alignat}
Again, the shifted coupling $f'_{LR}$ differs from the general
$f_{LR}$ in that now the equations of motion of Eq.\eqref{eq:eom} are
used to define a minimal operator basis.  The range of new physics
scales that can be probed in the Higgs sector may be extracted from
the parameters $\Delta$ collected in Table~\ref{tab:cplgs} which
  indicate potential deviations from the SM predictions of the Higgs
  couplings. According to the general analysis of independent
  operators introduced above, fermionic couplings and $gg,
  \gamma\gamma$ couplings prove particularly useful in this context
  while any deviations of $WW, ZZ$ couplings appear only on top of the
  large $hWW$ and $hZZ$ tree-level couplings. \medskip

\begin{table}[t]
\begin{tabular}{|l||c|c||c|c||c|}
\hline
$\Lambda_\ast$ [TeV]	 &  LHC  & HL-LHC & LC	 & HL-LC & HL-LHC + HL-LC \\
\hline\hline
$hWW$			 &  0.82 & 0.87   & 2.35 & 3.18  & 3.48 	  \\
$hZZ$			 &  0.74 & 0.87   & 2.75 & 3.48  & 3.89 	  \\
\hline
$htt$			 &  0.45 & 0.50   & 0.87 & 1.34  & 1.42 	  \\
$hbb$			 &  0.39 & 0.44   & 1.15 & 1.59  & 1.66 	  \\
$h\tau\tau$		 &  0.52 & 0.58   & 0.96 & 1.34  & 1.42 	  \\
\hline
$hgg$			 &  0.55 & 1.07   & 1.30 & 1.80  & 1.95 	  \\
$h\gamma\gamma$ 	 &  0.15 & 0.18   & 0.24 & 0.36  & 0.44 	  \\
\hline
\end{tabular}
\caption{Effective new physics scales $\Lambda_\ast$ extracted from the
	 Higgs coupling measurements collected in
	 Table~\ref{tab:cplgs}. The values for the
	 loop-induced couplings to gluons and photons contain only the
	 contribution of the contact terms, as the effects of the loop terms are
	 already disentangled at the level of the input values $\Delta$.}
\label{tab:Lambdas}
\end{table}
\begin{figure}[b!]
\includegraphics[width=0.6\textwidth]{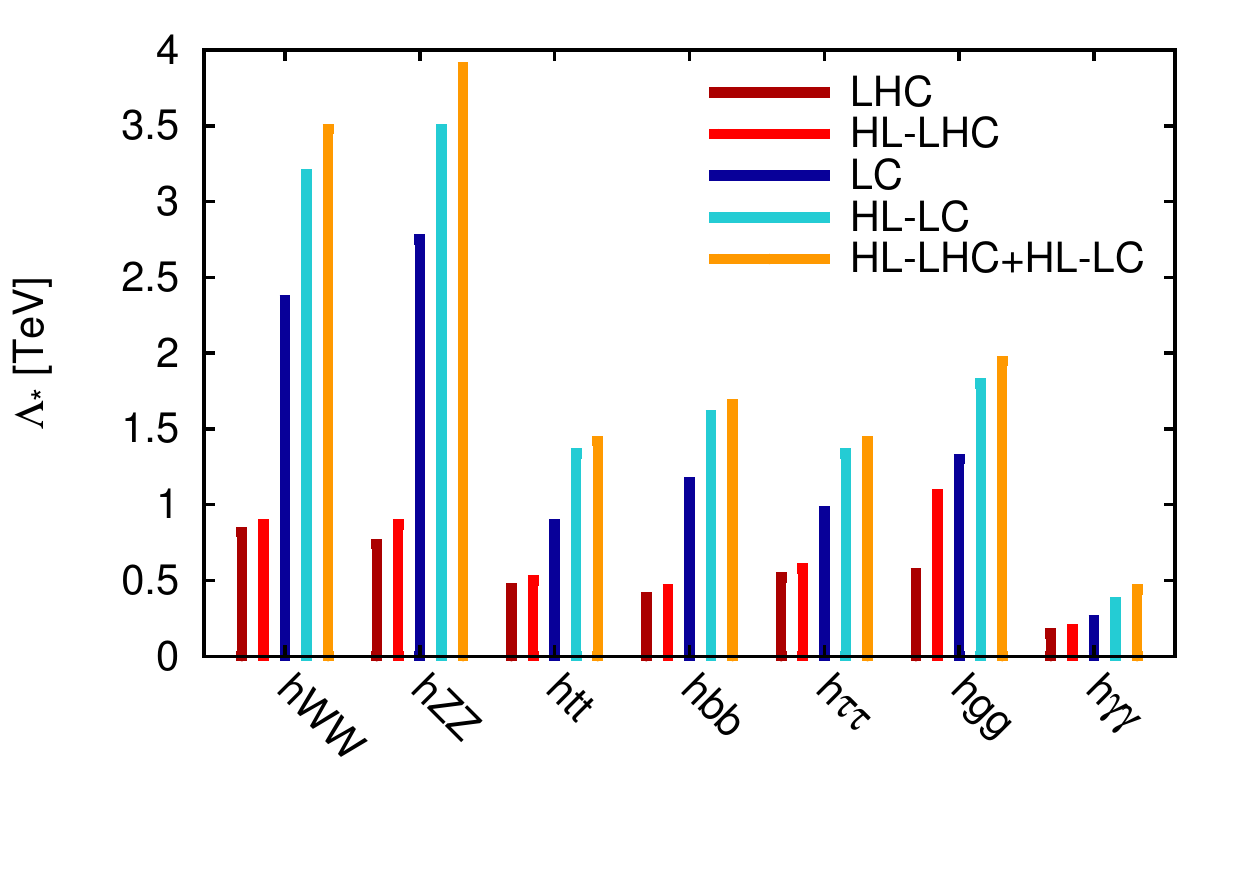}
\vspace{-1.2cm}
\caption{Effective new physics scales $\Lambda_\ast$ extracted from the
	 Higgs coupling measurements collected in
	 Table~\ref{tab:cplgs}. The values for the
	 loop-induced couplings to gluons and photons contain only the
	 contribution of the contact terms, as the effects of the loop terms are
	 already disentangled at the level of the input values $\Delta$.
(The ordering of
	 the columns from left to right corresponds to the legend
	 from up to down.)}
\label{fig:Lambdas}
\end{figure}

Now let us turn to the extraction of limits on the contributions of the $D=6$
operators. We again use SFitter~\cite{sfitter_higgs} for this purpose.
It is useful to define effective scales $\Lambda_\ast$ by factoring
out from the operators some typical coefficients,
like couplings of the kind $y,\,f'_X$ and loop factors $1/(16\pi^2)$.
In detail, we replace
\begin{alignat}{7}
\text{fermions} : & \qquad
&\Delta_f &= - \frac{v^2}{2\Lambda^2} \frac{f'_{LR}}{y}
	  &&\to \quad \frac{v^2}{2\Lambda_\ast^2 [f]}		    \notag \\
WW,ZZ :&
&\Delta_V &= -\frac{v^2}{2\Lambda^2} f'_{\phi 2}
	  &&\to \quad 2 \frac{v^2}{2\Lambda_\ast^2 [V_m]}	    \notag \\
gg :&
&\Delta_g &= - \frac{v^2}{2\Lambda^2} \frac{4 \cdot 16 \pi^2}{\zeta_g} f'_{GG}
	  &&\to \quad  \frac{4}{\zeta_g} \frac{v^2}{2\Lambda_\ast^2 [GG]}	     \notag \\
\gamma\gamma :&
&\Delta_\gamma &= - \frac{v^2}{2\Lambda^2} \frac{2 \cdot 16 \pi^2
}{\zeta_\gamma} \frac{f'_{BB} + f'_{WW}}{2}
	  &&\to \quad \frac1{\zeta_\gamma} \frac{v^2}{2\Lambda_\ast^2 [WW/BB]}	 \,,
\label{eq:coup_rescale}
\end{alignat}
where $GG$ denotes the gluonic contact term.
The factors $\zeta_g = A_{1/2}(4m_t^2/m_h^2) \simeq 4/3$ and
$\zeta_\gamma = (4/3) A_{1/2}( 4m_t^2/m_h^2) +
A_1( 4m_W^2/m_h^2) \simeq -6.5$ account for the total SM loop amplitude, see
Eqs.\eqref{eq:delaa} and~\eqref{eq:Afunc}.
In the effective Higgs--gluon and Higgs--photon couplings
the input values $\Delta$ already separate the contact terms from the
loop terms, induced by modified $htt$ and $hWW$ couplings. Therefore,
we can directly identify $\Delta_{g,\gamma}$ with the corresponding
contact terms without evaluating loop and contact terms
individually. While we only show the contribution of top and $W$ loops
in the formulae above, in the SFitter analysis all loop contributions
are properly taken into account.
The projected limits on the $\Lambda_\ast$ parameters as defined above
are collected in Table~\ref{tab:Lambdas} and Fig.~\ref{fig:Lambdas}.\medskip

As we can see, the effective new physics scales that can be probed in the
Higgs sector extend to a range from several hundred GeV to maximum
values beyond a TeV. However, bounds on new
particle masses exchanged at the Higgs vertex may be reduced significantly
by small couplings $M \sim \Lambda_\ast \sqrt{g^2/16\pi^2}$ as shown later
in this section. Thus, it depends on the specific model to what extent
precision Higgs analyses may explore high-mass domains in new physics
scenarios beyond direct searches at high-energy colliders.

\subsection{Strongly interacting Higgs field}
\label{sec:strongly1}

While originally light Higgs bosons were foreign to concepts of strong
electroweak symmetry breaking, the continuing support for light Higgs
bosons by electroweak precision analyses~\cite{ewfit} and finally the
LHC discovery of a light, narrow single Higgs boson~\cite{ATLASCMS}
suggested concepts within which a single light
state is embedded in a heavy strongly interacting sector.

An elegant formulation inspired by the AdS/CFT
correspondence~\cite{Maldacena:1997re} allows us to define a strongly
interacting Higgs sector in four space-time dimensions via a
Randall-Sundrum setup~\cite{Randall:1999ee} \textsl{i.e.} a slice of
five dimensional anti-de Sitter space bounded by two
3-branes~\cite{contino,Agashe:2004rs,Contino:2006qr}. The
5-dimensional picture can be used to investigate the dynamics of the
theory, which is not straightforwardly accessible in the strongly
coupled 4-dimensional picture~\cite{GroMMM}.  By using the standard
AdS/CFT dictionary (see {\sl e.g.}\ Ref.~\cite{Csaki:2005vy}) we can construct a low energy
effective theory that meets the phenomenologically observed symmetry
requirements, which can now be understood in a strongly interacting~\cite{ccwz}
large-$N$ conformal field theory (CFT) context.  The IR
brane-localized modes in the 5d picture correspond to additional
composite states that are indispensable for unitarity conservation in
the model, but can be neglected in a low-energy EFT approach.  The
relevant small parameter in the light Higgs effective theory is $v/f$,
where in the most strongly interacting setup the scale of the
additional states is $4 \pi f$.

Another way to model a strongly interacting Higgs sector with
  light Goldstone modes is by applying QCD-inspired chiral perturbation
  theory. In contrast to our discussion in Section~\ref{sec:dim6}
such a chiral Lagrangian is usually not written based on the linear
representation of the Higgs and Goldstone fields and hence does not
include an $SU(2)_L$ doublet $\phi$. The Higgs field then appears
as a singlet in the combination $h/v$, which gives more freedom to
define operators and does not allow for a one-to-one
correspondence of the power counting of the linear
Higgs operators and the non-linear chiral Lagrangian.
The non-linear and linear sets of
interaction operators are mutually
    equivalent concerning all possible Lorentz and $U(1)_{em}$ invariant
    couplings. If instead one only considers the leading components in
    each set, there is only partial
    correspondence between the leading operators~\cite{Brivio:2013pma,Contino:2013kra,buchalla}. \medskip

As a specific  example, where a one-to-one correspondence of leading
operators still holds,
   we show an ansatz for the higher-dimensional Lagrangian
based on the extra-dimensional strongly interacting theory described
above. Its leading terms have the same form as they would have in an
effective field theory based on $\phi^\dag \phi/f^2$ in a linear
representation. In the experimentally least vulnerable case where the
mass scale of the additional Kaluza--Klein states is given by $m_\rho
\sim 4 \pi f$ and we omit operators violating custodial symmetry, the
strongly interacting Higgs Lagrangian includes the
leading dimension-6 operators~\cite{silh}
\begin{alignat}{5}
{\cal L}_\text{SILH} &\supset
 \frac{c_h}{2f^2} \left[ \partial^\mu (h^\dagger h) \right]^2
 - \frac{c_6\lambda}{f^2} (h^\dagger h)^3
 + \sum_f \left( \frac{c_y y_f}{f^2} h^\dagger h {\bar f}_L hf_R + \text{h.c.} \right)
 + \ope \left( \frac{1}{(4 \pi f)^2} \right)
 + \ope \left( \frac{1}{(16 \pi^2 f)^2} \right) \; ,
\label{eq:silh}
\end{alignat}
where $c_h,c_6,c_y$ are numbers of order unity and
  $\lambda$, $y_f$ denote the Higgs self-coupling and the Yukawa
  couplings, respectively.\medskip

Even without using any equations of motions the structure of the
leading corrections to the Standard Model Higgs couplings is
relatively simple.  Deviations in the couplings to massive gauge
bosons are protected by custodial symmetry.
The leading deviations affect the Higgs
self-coupling and the Yukawa couplings, corresponding to the reduced
operator basis $\{\ope_{\phi 2}, \ope_{\phi 3}, \ope_{LR}\}$ as
compared to the basis defined in Eq.\eqref{eq:d6_basis}.
In addition, the first term in Eq.\eqref{eq:silh} generates a
Higgs wave function renormalization, which leads to a universal correction
of all Higgs couplings as $1+\Delta = (1-c_hv^2/f^2)^{1/2}$.
The phenomenologically relevant gauge boson and
fermion couplings are modified by the two parameters $\xi$ and
$\xi c_y/c_h$~\cite{Gupta:2012mi},
where
\begin{equation}
\xi = c_h \Bigl (\frac{v}{f}\Bigr)^2
\end{equation}
is related to the Goldstone scale $f$ relative to the
standard Higgs vacuum expectation value $v$.
Compared to the weakly interacting models discussed
before, this two-parameter setup corresponds to the simplest 2HDM
scenarios~\cite{dcpl2}. \medskip

\begin{table}[t]
\begin{tabular}{|l||c|c||c|c||c|}
\hline
$\xi$		& LHC	 & HL-LHC & LC	   & HL-LC & HL-LHC+HL-LC \\
\hline
universal	& 0.076  & 0.051  & 0.008  & 0.0052 & 0.0052	  \\
non-universal	& 0.068  & 0.015  & 0.0023 & 0.0019 & 0.0019	  \\
\hline \hline
$f$ [TeV]	&	 &	  &	   &	    &		  \\
\hline
universal	& 0.89	 & 1.09   & 2.82   & 3.41   & 3.41	  \\
non-universal	& 0.94	 & 1.98   & 5.13   & 5.65   & 5.65	  \\
\hline
\end{tabular}
\caption{Estimates of the parameter $\xi = (v/f)^2$ and the Goldstone
	     scale $f$ for various experimental set-ups and two
	     different fermion embeddings (universal, non-universal).}
\label{tab:xitab}
\end{table}

The ratio $c_y/c_h$ can be predicted
in the context of holographic
Higgs Models, in which strongly coupled theories in four dimensions
are identified with weakly coupled theories in five dimensions.
In theories in which the global symmetry
SO(5) is broken to SO(4), the Standard Model fermions may be assigned
either to spinorial or fundamental SO(5) representations, changing
the Higgs couplings either universally ($c_y/c_h=0$) or separately ($c_y/c_h \neq
0$) for Standard Model vectors and
fermions.
The spinorial case where all Higgs couplings are suppressed universally by a
factor $(1-\xi)^{1/2}$~\cite{Agashe:2004rs} is covered by
the analysis of portal models. \medskip

In a closely related scenario~\cite{Contino:2006qr}
universality is broken to the extent that the Higgs coupling
of vector particles is reduced by the standard coefficient,
but the coupling of fermions by a different coefficient,
\begin{equation}
1 + \Delta_V = \sqrt{1-\xi} \approx 1 - \frac{\xi}{2}
\qquad	\qquad \qquad
1 + \Delta_f = \frac{1-2\xi}{\sqrt{1-\xi}} \approx 1- 3 \frac{\xi}{2}  \,.
\label{eq:xinonunivf}
\end{equation}
for $\xi \ll 1$.
Based on the estimates of potential deviations from SM Higgs couplings,
bounds on the parameter $\xi$ and the ensuing scale $f$ are presented
in Table~\ref{tab:xitab}. \medskip

The typical bounds on $\xi$ range from 7.6\% at the LHC
up to 2 permille for the non-universal scenario at an upgraded linear
collider. The expectations in the non-universal scenario are
stronger than those in the universal scenario, apparent from the factor 3
in the expanded version of $\Delta_f$ in Eq.~\eqref{eq:xinonunivf}
compared with $\Delta_V$ in both scenarios. In terms of the
Goldstone scale $f$ this corresponds to scales between just
under a TeV to more than 5~TeV. These numbers exceed the
limits from electroweak precision data, which yield $f \gesim 750$~GeV~\cite{sew}.
However, no meaningful limit can be obtained from LHC
Higgs data if we relax the assumption that there are no non-SM
Higgs decay modes~\cite{Gupta:2012mi,Contino:2010mh}.

\section{Mixing effects}
\label{sec:mix}

Mixing phenomena are a general consequence of multi-field structures
in the scalar sector.  Assuming that one of the states is essentially
identical with the SM Higgs state, mixing nevertheless affects masses
and couplings, inducing potentially small deviations from the SM
values for large scales of the new physics sector. These mixing effects can
in principle be quite complex. We will discuss in detail
three interesting examples which illustrate the basic features.

\subsection{Higgs portal}

The large dark component of matter in the Universe strongly suggests a
dark sector with potentially complex structure~\cite{mambrini}. This
sector may interact with the Standard-Model sector through the Higgs
portal~\cite{Bij,Wells1,Bock,Wells2}, the two sectors
coupled by a renormalizable quartic interaction,
\begin{equation}
   \lag_p = - \eta \, |\phi_s|^2 |\phi_d|^2
\end{equation}
between the SM Higgs field $\phi_s$ and a corresponding dark Higgs
field $\phi_d$. The individual interactions conform with the
standard quartic interactions of spontaneous symmetry breaking with
strengths $\lambda_s$ and $\lambda_d$,
\begin{equation}
   {\lag_4} = - \frac{\lambda_s}{2} \, |\phi_s|^4 - \frac{\lambda_d}{2} \,
|\phi_d|^4 \, .
\end{equation}
Before symmetry breaking the theory is $SU(3) \times SU(2) \times U(1)$
gauge-invariant.  After symmetry breaking in the SM and hidden sectors
the system is described by the mass matrix~\cite{Bock}
\begin{equation}
   \mathcal{M}^2 = \left( \begin{array}{cc}   \lambda_s v^2_s  &    \eta v_s v_d    \\
						\eta v_s v_d	 &  \lambda_d v^2_d
			    \end{array} \right) \,,
\end{equation}
where $v_s \equiv v^\text{SM}$ and $v_d$ are the vacuum expectation
values of the Higgs fields of the coupled system, the coupling mediated
by the mixed term $\eta$.  \medskip

The mass spectrum of the two Higgs bosons,
\begin{alignat}{7}
   m^2_{s_1} &\simeq \lambda_s v^2_s - \eta^2 v^2_s / \lambda_d  \notag \\
   m^2_{d_1} &\simeq \lambda_d v^2_d + \eta^2 v^2_s / \lambda_d
\end{alignat}
splits characteristically into a light SM-type state and a new heavy
state. The initial masses in the two sectors are pulled apart by the
mutual interaction, inducing a mass splitting of the order $v_s$
times the interaction strength $\eta$.	Thus, the mixing effect
on the mass spectrum is determined by the light SM scale and not by the
heavy scale, and a sufficiently small mixing parameter $\eta$ allows the light
system to approach the structure of the Standard Model. \medskip

\begin{figure}[t]
\includegraphics[width=0.44\textwidth]{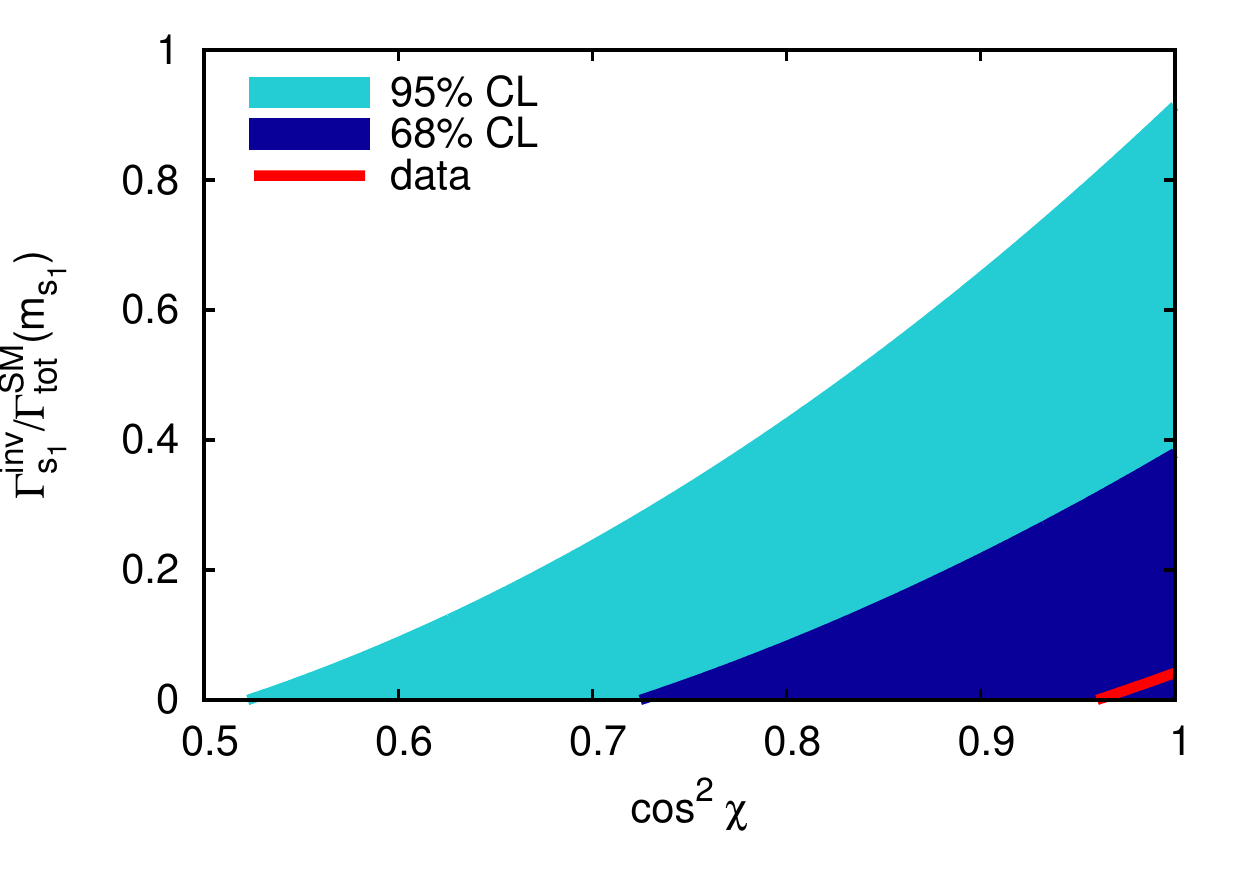} \hspace{5mm}
\includegraphics[width=0.44\textwidth]{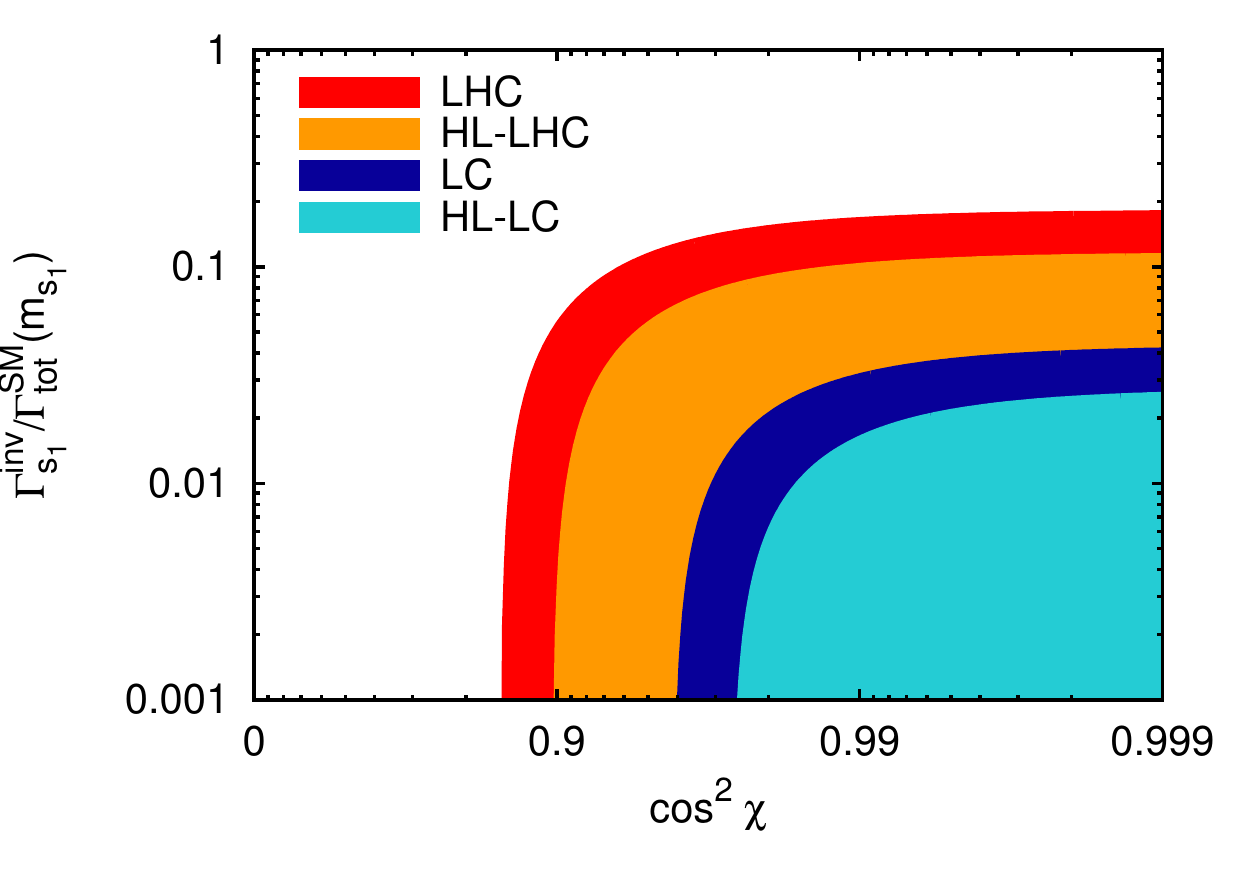}
\vspace{-1.2em}
\caption{Best-fit contours for the mixing parameter $\cos^2\chi$ and
  the hidden decay width $\Gamma^\text{inv}$ in the Higgs portal
  model. Left: Shown is the relation originating from the best-fit
  point of the Higgs couplings and the 68\% and 95\% C.L.\ error bars,
  using the result of Ref.~\cite{dcpl2}, which includes LHC data up to
  around Moriond/Aspen 2013.  Right: Expected improvements at the 95\%
  C.L.\ from future running of LHC, HL-LHC, LC and HL-LC.}
\label{fig:MRauch}
\end{figure}

The two mass eigen-fields $s_1$, $d_1$ are rotated out of the current
fields $s,d$ by
\begin{alignat}{7}
   s_1 &= + \cos\chi\, s + \sin\chi\, d  \notag  \\
   d_1 &= - \sin\chi\, s + \cos\chi\, d  \, ,
\end{alignat}
with the mixing angle $\chi$ given by
\begin{equation}
   \tan (2\chi) \simeq - \frac{2 \eta\, v_s }{\lambda_d  v_d}  \,.
\end{equation}
This size of the phenomenological mixing angle is determined by $\eta$
and, in contrast to the mass spectrum, by the ratio $v_s / v_d$ of the SM
scale over the high scale, as naively expected.
The mixing affects all the couplings of the SM-like Higgs boson
universally,
\begin{equation}
g_{s_1} = \cos\chi\; g^\text{SM}_h  \,,
\label{eq:portalcpl}
\end{equation}
which is certainly the easiest way to quantify large deviations
from the Standard Model in existing experimental data. The present
bound on $\cos^2 \chi$ is shown in the left panel of Fig.~\ref{fig:MRauch}. Despite the
fact that the mass scale $v_d$ is much larger than $v_s$, there could
still be light particles in the dark sector, just like all SM fermions
but the top quark have a mass much smaller than $v$. This opens up the
possibility for invisible decays of the SM-like state $s_1$. Therefore, we combine it
with the estimate of the partial width for invisible Higgs decay
channels.
The improvements foreseen from LHC, HL-LHC, LC
and HL-LC in the coming years and later in the future are displayed in
Fig.~\ref{fig:MRauch} (right), reinterpreting the results given in
Table~\ref{tab:cplgs}.	It is apparent that a fine-grain picture of the
Higgs boson can be drawn by analyzing the couplings. \medskip

Denoting $\delta_\chi = \sqrt{1 + \tan^2 (2\chi)} - 1$,
which reduces for small mixing to $\delta_\chi \simeq 2 \chi^2$,
the individual Higgs vacuum expectation values are shifted by
\begin{alignat}{7}
   \lambda_s  v^2_s &= m^2_{s_1} + \frac{1}{2} m^2_{d_1}\, \delta_\chi \notag \\
   \lambda_d  v^2_d &= m^2_{d_1} - \frac{1}{2} m^2_{d_1}\, \delta_\chi
\end{alignat}
for $m_{d_1} \gg m_{s_1}$. Evidently, the measurement of the light
Higgs mass $m_{s_1}$ and the mixing parameter $\delta_\chi$ gives rise
to an upper limit on the heavy Higgs mass $m_{d_1}$
\begin{equation}
   m_{d_1}^2 \leq \frac{2 m_{s_1}^2}{\delta_\chi} \,.
\label{eq:portallimit}
\end{equation}
As naturally expected in quantum mechanics, the masses approach each
other for large mixing while the gap spreads for small mixing. \medskip

\begin{figure}[t]
\includegraphics[width=0.44\textwidth]{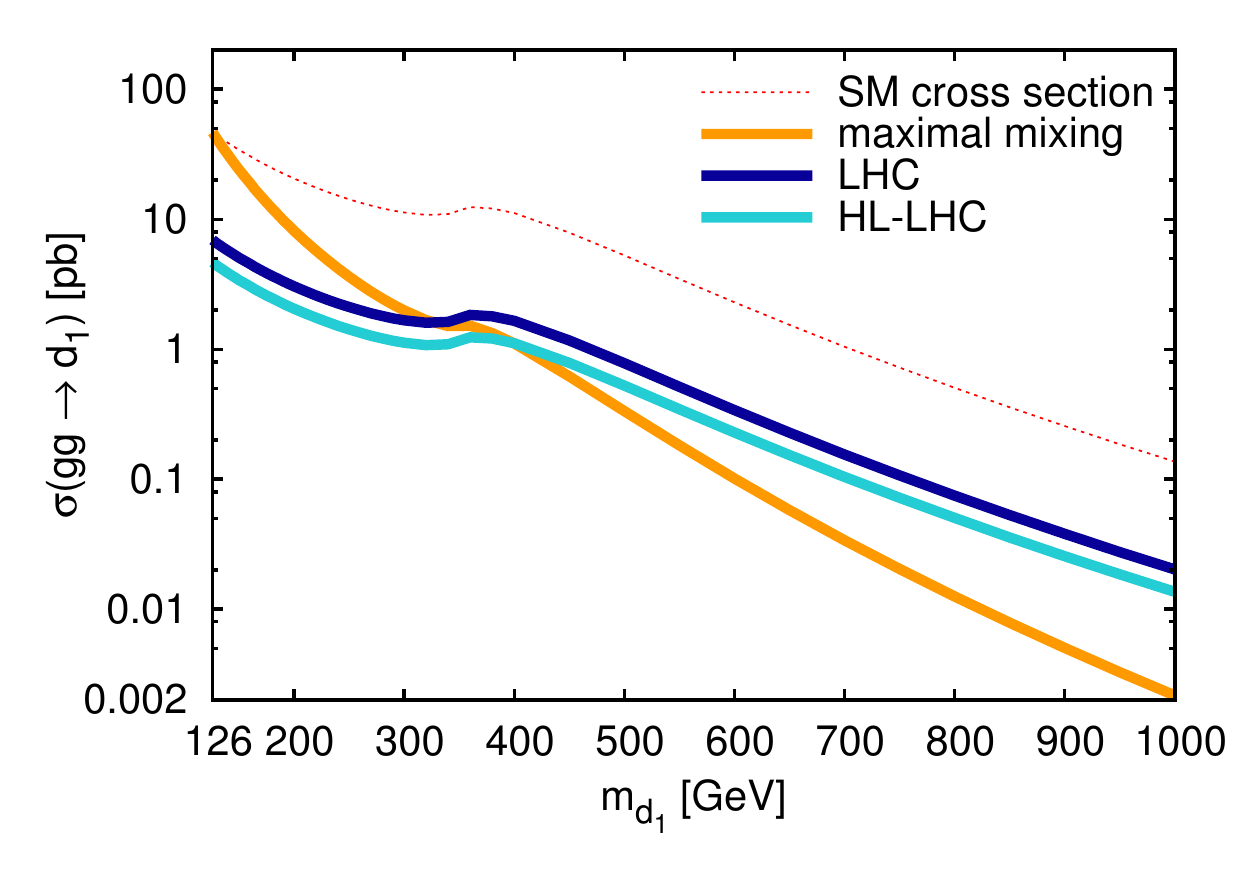} \hspace{5mm}
\includegraphics[width=0.44\textwidth]{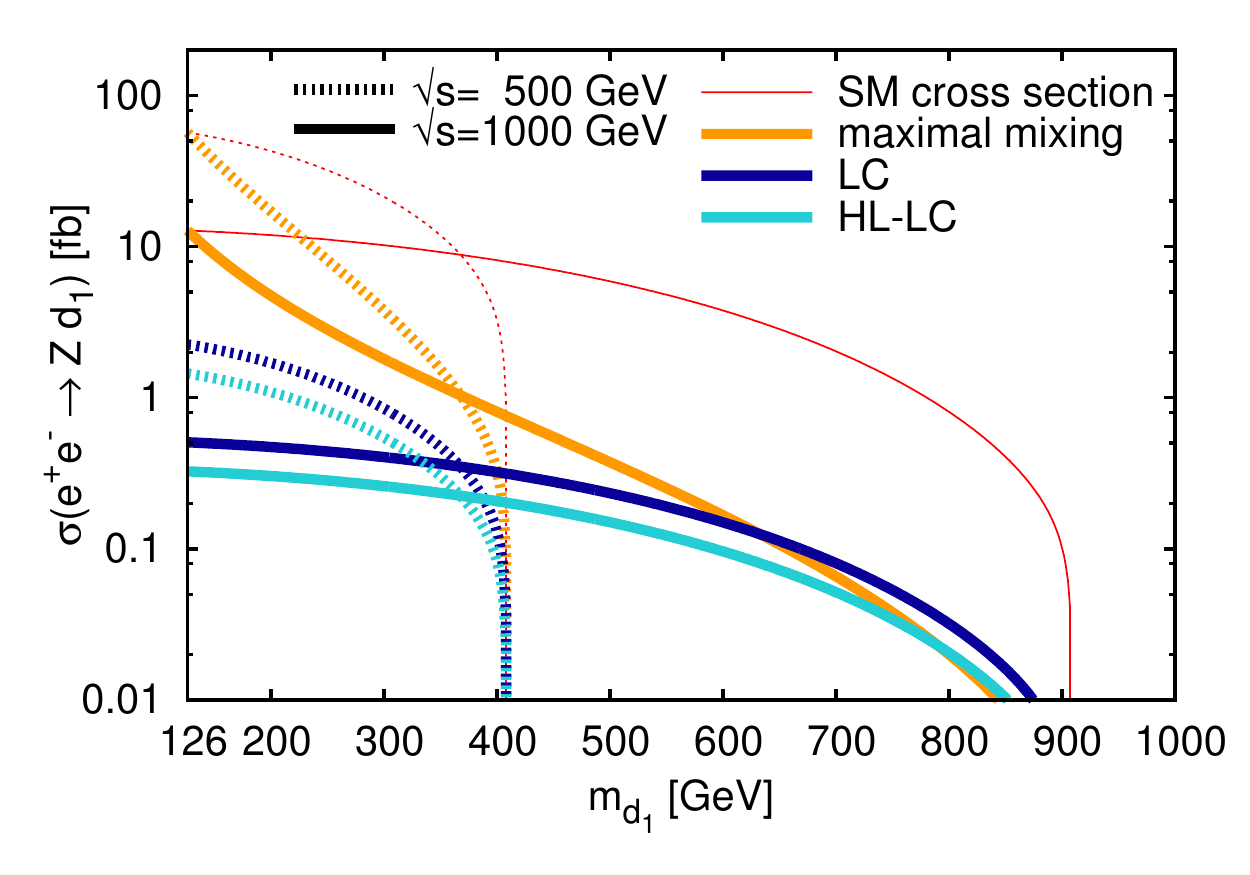}
\vspace{-1.2em}
\caption{Production cross sections of the heavy Higgs boson in portal
  models, as a function of the Higgs mass (left) at the LHC;
  (right) at the LC. The yellow curve shows the $d_1$
  production cross section for
  maximal allowed mixing according to Eq.\eqref{eq:portallimit},
  while the blue curves correspond to the maximal allowed mixing if
  the $s_1$ couplings agree with the Standard Model within 95\% C.L.\ after the
  standard and high-luminosity run of the respective collider, see
  Eq.\eqref{eq:portalcpl}.}
\label{fig:portal}
\end{figure}

The interplay between heavy Higgs masses and mixings in observing this
state either at LHC or LC is demonstrated in
Fig.~\ref{fig:portal}. As a function of the
Higgs mass, the mixing parameters are chosen according to three
different scenarios. These are the maximal allowed mixings given by
Eq.\eqref{eq:portallimit} and assuming that the $s_1$ couplings agree with the Standard Model at
the 95\% C.L.\ after both the standard and high-luminosity run of the
respective collider.
For the LHC we show the gluon fusion Higgs
production cross sections including NNLO and NNLL QCD
corrections~\cite{LHCHXSWG}, while for the LC the Higgs-strahlung
process $e^+ e^- \rightarrow Z + h$ is presented with cross sections of 67.1 and
13.4 fb for a SM Higgs boson with a mass of 126 GeV at LC facilities
of energies 500 GeV and 1 TeV~\cite{ilc}, respectively.

\subsection{Two Higgs doublets and the MSSM}
\label{sec:2hdm}

In a general Two-Higgs-Doublet model (2HDM), the physical
states are mixtures of the components of the two doublets $\phi_1$ and
$\phi_2$ \cite{2hdm,dcpl}.
The scalar potential can be written as
\begin{alignat}{7}
V = &\; m_{11} |\phi_1|^2 + m_{22}^2 |\phi_2|^2 - m_{12}^2(\phi_1^\dagger \phi_2 +
\text{h.c}) + \lambda_1 |\phi_1|^4 + \lambda_2 |\phi_2|^4 \nonumber \\
&+ \lambda_3 |\phi_1|^2 |\phi_2|^2 + \lambda_4 |\phi_1^\dagger \phi_2|^2
 + \frac{1}{2}\lambda_5 [(\phi_1^\dagger \phi_2)^2 + \text{h.c}]\,.
\label{eq:2hdm_pot}
\end{alignat}
The Higgs--fermion couplings depend on the specific type of the 2HDM.
To ensure natural suppression of flavor-changing neutral currents, one usually
demands that one type of fermions couples only to one Higgs doublet.
This pattern can be imposed by a
global $\mathbb{Z}_2$ discrete symmetry, under which
$\phi_{1,2} \to \mp \phi_{1,2}$, and which has been assumed in the potential
Eq.\eqref{eq:2hdm_pot}. So all terms in Eq.\eqref{eq:2hdm_pot} include
an even power of each of the Higgs fields $\phi_1$ and $\phi_2$. Both Higgs fields
acquire vacuum expectations values,
$v_1$ and $v_2$, with $v^2_1+v^2_2=v^2$ and $\tan\beta = v_2/v_1$.
Depending on the $\mathbb{Z}_2$ charge assignments, the following four cases of
coupling the Higgs doublets to fermions are possible~\cite{Barger:1989fj}:
\begin{itemize}
\item \underline{type I:} all fermions couple only to $\phi_2$;
\item \underline{type II:} up-/down-type fermions couple to
$\phi_2$/$\phi_1$, respectively;
\item \underline{lepton-specific:} quarks couple to $\phi_2$ and charged
leptons couple to $\phi_1$;
\item \underline{flipped:} up-type quarks and leptons couple to $\phi_2$ and
down-type quarks couple to $\phi_1$.
\end{itemize}
After electroweak symmetry breaking the Higgs sector consists of three
neutral Higgs bosons, two CP-even ones $h^0, H^0$ and a CP-odd one $A^0$,
as well as of two charged Higgs bosons $H^\pm$. Leaving aside
modifications of the loop decays from the individual loop particle contributions and
ignoring Higgs-to-Higgs decays, which alter the branching ratios, the
partial widths of the light scalar $h^0$ are modified relative to the Standard Model
through mixing effects of the two Higgs doublets. They can be expressed in terms
of $\beta=\arctan v_2/v_1$ and $\alpha$, the mixing angle between the two
CP-even Higgs bosons $h_0$ and $H_0$, see Table~\ref{tab:thdmcoupl}.

\begin{table}[t]
\begin{tabular}{|c|cccc|}
\hline
$\;\dfrac{\Gamma_\text{2HDM}[h^0\to X]}{\Gamma_\text{SM}[h\to X]}\;$
 & type I & type II & lepton-spec. & flipped \\
\hline
$VV^\ast$
 & $\sin^2(\beta-\alpha)$ & $\sin^2(\beta-\alpha)$
 & $\sin^2(\beta-\alpha)$ & $\sin^2(\beta-\alpha)$ \\[.5ex]
$\bar{u}u$
 & $\dfrac{\cos^2\alpha}{\sin^2\beta}$ & $\dfrac{\cos^2\alpha}{\sin^2\beta}$
 & $\dfrac{\cos^2\alpha}{\sin^2\beta}$ & $\dfrac{\cos^2\alpha}{\sin^2\beta}$
\\[2ex]
$\bar{d}d$
 & $\dfrac{\cos^2\alpha}{\sin^2\beta}$ & $\dfrac{\sin^2\alpha}{\cos^2\beta}$
 & $\dfrac{\cos^2\alpha}{\sin^2\beta}$ & $\dfrac{\sin^2\alpha}{\cos^2\beta}$
\\[2ex]
$\ell^+\ell^-$
 & $\dfrac{\cos^2\alpha}{\sin^2\beta}$ & $\dfrac{\sin^2\alpha}{\cos^2\beta}$
 & $\dfrac{\sin^2\alpha}{\cos^2\beta}$ & $\dfrac{\cos^2\alpha}{\sin^2\beta}$
\\[.5ex]
\hline
\end{tabular}
\caption{Partial widths of the light Higgs boson $h^0$ in different realizations
of the 2HDM, relative to the SM widths at leading order.}
\label{tab:thdmcoupl}
\end{table}

\begin{figure}[b]
\makebox[0.23\textwidth]{\centering type-I}
\makebox[0.23\textwidth]{\centering type-II}
\makebox[0.23\textwidth]{\centering lepton-specific}
\makebox[0.23\textwidth]{\centering flipped}\\[-1em]
\mbox{
\raisebox{-\height}{\includegraphics[width=0.23\textwidth]{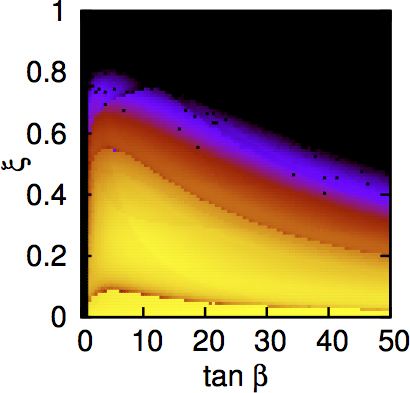}}
\raisebox{-\height}{\includegraphics[width=0.23\textwidth]{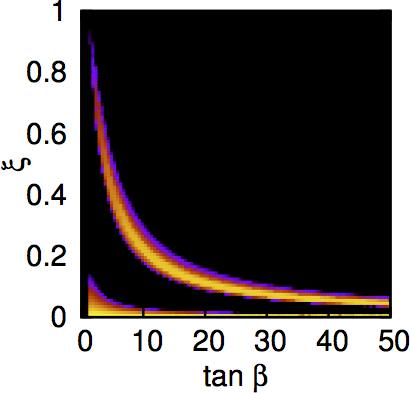}}
\raisebox{-\height}{\includegraphics[width=0.23\textwidth]{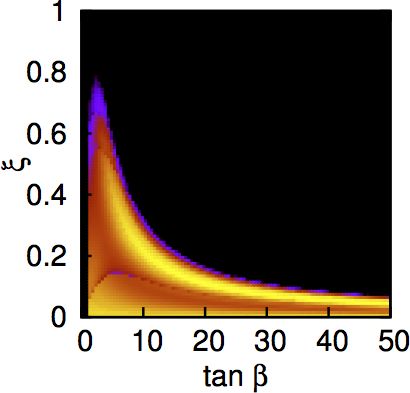}}
\raisebox{-\height}{\includegraphics[width=0.23\textwidth]{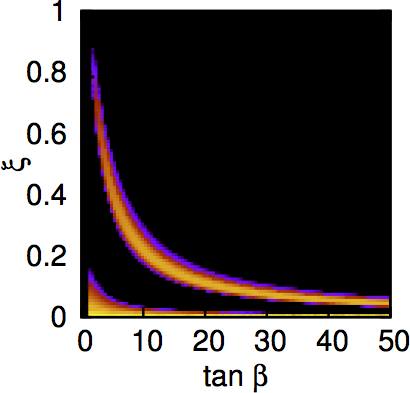}}
\raisebox{-\height}{\includegraphics[width=0.035\textwidth]{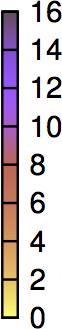}}}
\caption{Allowed ranges for the modification parameter $\xi$ in a 2HDM in the
decoupling limit, based on data from Ref.~\cite{ATLASCMS2}. The plots show the
correlated relative log-likelihood $-2\,\Delta(\log\cal L)$ as a function of
$\tan\beta$, from Ref.~\cite{dcpl2}.}
\label{fig:2HDM}
\end{figure}

If the heavy Higgs bosons ($H^0$, $A^0$, and $H^\pm$) have
masses much larger than $v$, one enters a regime where the physics of the light
Higgs $h^0$ can be described by an effective theory~\cite{dcpl,dcpl2}. In this case,
commonly known as the decoupling limit, the heavy Higgs masses are
approximately given by
\begin{equation}
   m_{A^0}^2,\, m_{H^0}^2,\, m_{H^\pm}^2 = \frac{2m^2_{12}}{\sin (2\beta)} + \ope(v^2) \,,
\end{equation}
while the properties of the lightest CP-even Higgs boson $h^0$ are
close to the Standard Model.  The leading modification of the partial
widths of $h^0$ in relation to the Standard Model can be expressed as
an expansion in the parameter
\begin{equation}
\xi = \frac{v^2}{2 m_{A^0}^2} \; \sin^2(2\beta) \;
\left[
 \lambda_1-\lambda_2+
(\lambda_1+\lambda_2-\lambda_3-\lambda_4-\lambda_5)\cos 2\beta \right] \,.
\end{equation}
For the factors in Table~\ref{tab:thdmcoupl} one thus finds
\begin{align}
\sin^2(\beta-\alpha) &\approx 1-\xi^2, &
\frac{\cos^2\alpha}{\sin^2\beta} &\approx 1+2\xi\cot\beta, &
\frac{\sin^2\alpha}{\cos^2\beta} &\approx 1-2\xi\tan\beta.
\label{eq:thdm_fact}
\end{align}
Numerically, for $\lambda_i \sim \ope(1)$ and $\tan\beta \approx 1$,
the parameter $\xi \approx 0.03/(m_{A^0}/\text{TeV})^2$, so one expects
corrections of tens of percent for moderate values of $m_{A^0}$ and
$\tan\beta$. The shape of the decoupling limit in the different model
setups can be seen in Fig.~\ref{fig:2HDM}. The preference for type-I
models in a comparably wide parameter range is that it separates Higgs
couplings to gauge bosons and fermions and makes it easy to accommodate
the slightly enhanced $H \to \tau \tau$ rate.  Many dedicated 2HDM
analyses based on the Higgs couplings measured at the LHC are
available for the different model setups~\cite{2hdm_fits}. Given the
generic size of experimental error bars and the fact that after the
recent experimental updates all channels are in broad agreement with
the Standard Model predictions, none of them shows a clear sign for
such mixing effects.\medskip

Minimal Supersymmetric Models (MSSM) form a subgroup
of the general 2HDM type-II. The quartic couplings of the 2HDM scalar
potential are restricted to special values given by the $SU(2)_L$ and
$U(1)$ gauge couplings $g$ and $g'$,
\begin{equation}
\lambda_1 = \lambda_2 = -\dfrac{1}{2}\lambda_3 = \dfrac{1}{8}(g^2+g'^2),
\qquad
\lambda_4 = -\lambda_5 = -g^2 \; .
\end{equation}
Because just like in the Standard Model the Higgs masses are determined
by the quartic couplings, this structure predicts the maximum mass of
the lightest Higgs scalar~\cite{mssm_mh}.
In the decoupling limit, {\sl i.e.}~the limit of heavy
Higgs bosons $H^0, A^0$ and $H^\pm$, the partial widths of this SM-like
light state scale as
\begin{alignat}{7}
\frac{\Gamma_\text{SUSY}[h^0\to VV^\ast]}{\Gamma_\text{SM}[h\to VV^\ast]}
 &\approx 1- \frac{m_Z^4\sin^2 2\beta}{m_{A^0}^4} \; (\cos 2\beta + R_t)^2\,,
\notag \\
\frac{\Gamma_\text{SUSY}[h^0\to uu]}{\Gamma_\text{SM}[h\to uu]}\;
 &\approx 1+ \frac{4m_Z^2\cos^2\beta}{m_{A^0}^2} \; (\cos 2\beta + R_t) \,,
\notag \\
\frac{\Gamma_\text{SUSY}[h^0\to dd]}{\Gamma_\text{SM}[h\to dd]}\,\;
 &\approx 1- \frac{4m_Z^2\sin^2\beta}{m_{A^0}^2} \; (\cos 2\beta + R_t)\,,
\end{alignat}
where $R_t$ captures the SUSY radiative corrections, which are
dominated by the top/stop loop contributions:
\begin{alignat}{5}
R_t \approx \frac{3(g^2+g'^2)}{16\pi^2\sin^2\beta} \,\frac{m_t^4}{m_Z^4}
 \biggl [ &\log\frac{m_{\tilde{t}_1}m_{\tilde{t}_2}}{m_t^2} +
 (A_t - \mu \cot 2\beta)\frac{A_t-\mu\cot\beta}{m_{\tilde{t}_1}^2 - m_{\tilde{t}_2}^2}
 \log\frac{m_{\tilde{t}_1}^2}{m_{\tilde{t}_2}^2} \notag \\
 & + (A_t^2-\mu^2 - 2A_t\mu\cot 2\beta)\biggl ( \frac{A_t-\mu\cot\beta}{m_{\tilde{t}_1}^2 - m_{\tilde{t}_2}^2}
      \biggr )^2 \biggl ( 1- \frac{m_{\tilde{t}_1}^2 +
      m_{\tilde{t}_2}^2}{m_{\tilde{t}_1}^2 - m_{\tilde{t}_2}^2}
      \log\frac{m_{\tilde{t}_1}}{m_{\tilde{t}_2}}\biggr )
 \biggr ]\,,
\end{alignat}
where $\mu$ denotes the higgsino mass parameter,
$m_{\tilde{t}_{1,2}}$ the stop masses and $A_t$ the soft SUSY breaking
trilinear coupling of the stop sector.
To achieve compatibility with the relatively `large' Higgs boson mass
of $m_h \simeq 126$~GeV in the MSSM, rather high scales
$m_{\tilde{t}_{{1,2}}}
\sim \ope(1\text{ TeV})$ for the stop masses help, at the same time
enhancing the corrections in $R_t$.\medskip

\begin{figure}
\includegraphics[width=0.5\textwidth]{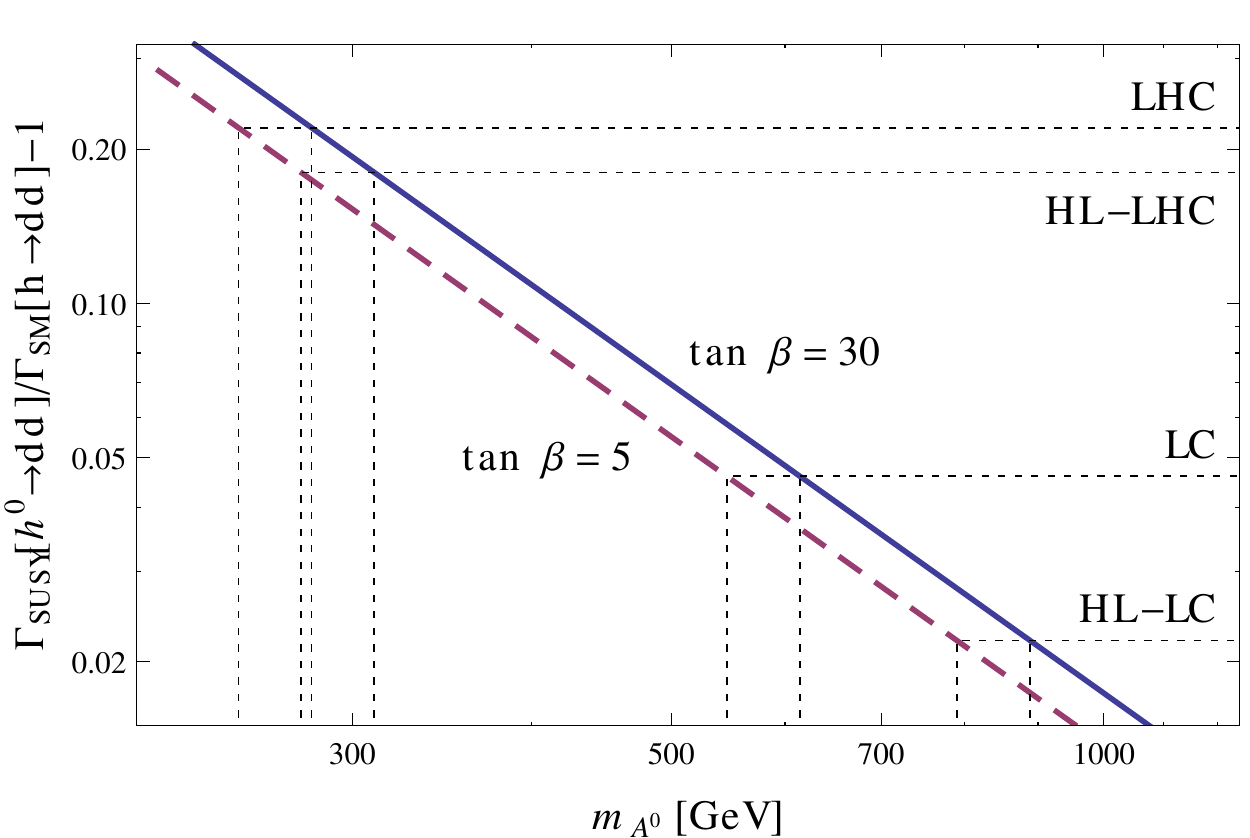}
\vspace{-1ex}
\caption{Dependence of the Higgs tree-level decay width into down-type fermions on $m_{A^0}$
within the MSSM, for two examples of $\tan\beta = 5$ and
30, as well as $m_{\tilde{t}_1}m_{\tilde{t}_2} = 1$~TeV$^2$ and $A_t-\mu\cot\beta
\ll m_{\tilde{t}_i}$. Also shown are the expected limits from the LHC,
HL--LHC, LC, HL--LC.}
\label{fig:MSSM}
\end{figure}

Due to the small number of free parameters in the tree-level Higgs
sector of the MSSM, as compared to the 2HDM, one can extract significant
information on the supersymmetric parameter space from LHC measurements.
A strong constraint comes from the light Higgs mass, but additional information
is obtained from the measured Higgs couplings in such a parameter
study~\cite{Carena:2001bg,mssm_coup_fits,marcela_old}.
In Fig.~\ref{fig:MSSM} we show how the CP-odd mass $m_{A^0}$ can be
estimated from measurements of the light Higgs
couplings. This is illustrated for two
representative values $\tan\beta = 5$ and 30.
For large values of $\tan\beta$ we need to keep in mind that vertex
loop corrections can become equally
important as the tree-level mixing effects, see
Section~\ref{sec:stb}.

\subsection{Next-to-Minimal Supersymmetric Standard Model (NMSSM)}

\noindent
In the NMSSM \cite{nmssm1,nmssm2,nmssm3} an additional singlet
superfield $S$ is added to the MSSM spectrum. The $\mu$ parameter is
generated dynamically through the coupling of $S$ to the two Higgs
doublet fields, $\lambda S H_u H_d$.  In order to avoid a massless
axion, the Peccei-Quinn symmetry \cite{pqsymmetry} is broken by the
introduction of a cubic coupling of the singlet field $S$, $\kappa
S^3/3$, in the scale-invariant superpotential. Furthermore, the MSSM
soft SUSY breaking  Lagrangian is extended by a soft SUSY breaking mass
term and trilinear soft SUSY breaking interactions for the singlet
field,
\begin{eqnarray}
- {\cal L}_{S,\text{soft}} = m_S^2 |S|^2 + \lambda A_\lambda H_u H_d S
+ \frac{1}{3} \kappa A_\kappa S^3 \;.
\end{eqnarray}
After electroweak symmetry breaking the Higgs sector consists of seven Higgs bosons, three CP-even states $H_i$ ($i=1,2,3$), two CP-odd
bosons $A_1$ and $A_2$,
and two charged Higgs bosons $H^\pm$. The
CP-even/odd states are ordered by ascending mass. Depending on the
choice of parameters, either the lightest or the next-to-lightest
Higgs boson can be the SM-like Higgs $h$. Its upper mass bound is
given by
\begin{eqnarray}
m_h^2 \approx M_Z^2 \cos^2 2\beta + \frac{\lambda^2 v^2}{2} \sin^2
2\beta + \Delta m_h^2 \;,
\end{eqnarray}
where $v\approx 246$~GeV. The tree-level contribution is maximized for small
values of $\tan\beta$, and due to the additional term proportional to
$\lambda$ a radiative correction $\Delta m_h^2$ of only $\sim (75$~GeV$)^2$
is required to achieve $m_h \approx 126$~GeV, for $\lambda = 0.6$ and
$\tan\beta=2$. The CP-even/CP-odd Higgs mass eigenstates are
admixtures of the singlet components $h_s/a_s$ and the doublet
components $h_u,h_d/a_u,a_d$, leading to suppressed couplings to the
SM particles due to the singlet component. \\

In this section we will first summarize the evidence that the
experimental observation of the 126-GeV Higgs state and the measurement
of its couplings are compatible with the NMSSM. And second,
that sum rules for precisely measured couplings can be exploited to prove
that observing three neutral MSSM-type Higgs bosons does not close the system
but may point to [two] more heavy neutral particles as predicted in the NMSSM. \\

\begin{figure}
\includegraphics[width=0.47\textwidth]{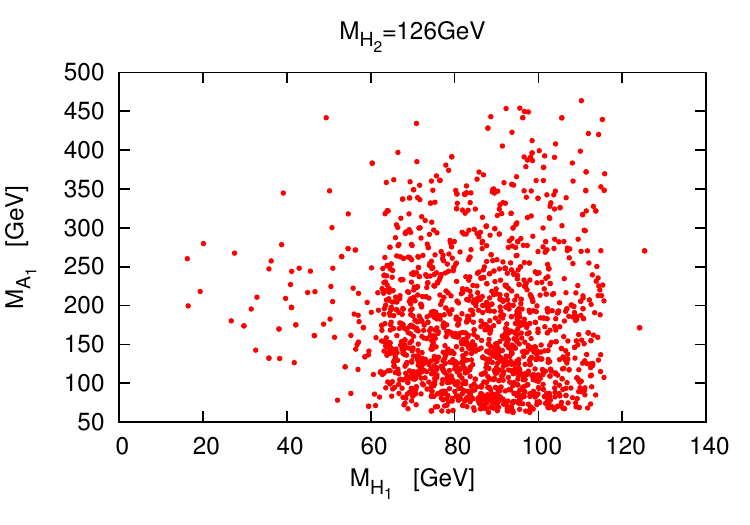}
\includegraphics[width=0.47\textwidth]{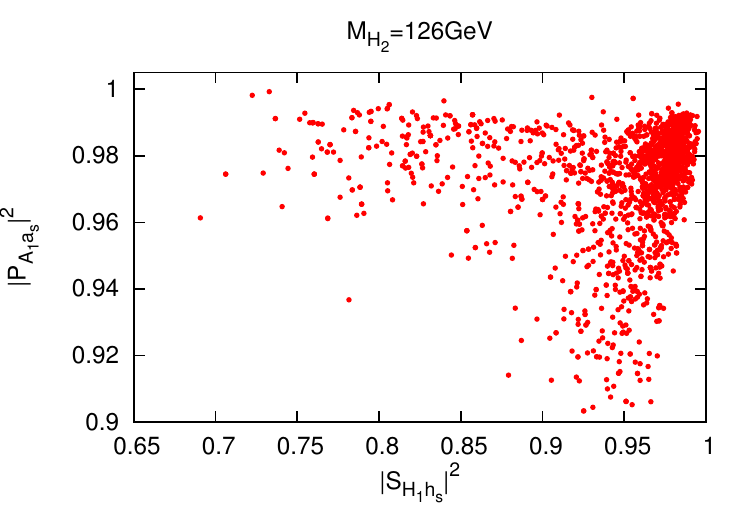}
\vspace{-1.2em}
\caption{Left: Mass values of $A_1$ and $H_1$ in GeV. Right: The
  singlet admixtures $|P_{A_1 a_s}|^2$ and $|S_{H_1h_s}|^2$ of the
  lightest pseudoscalar $A_1$ and of the scalar $H_1$.}
\label{fig:MMM1}
\end{figure}

The present LHC Higgs search results \cite{ATLASCMS} can be accommodated
in the NMSSM Higgs sector \cite{nmssmscans,furtherscans}. With one of
the CP-even
Higgs bosons being SM-like, the LEP constraints can be avoided for the
light CP-even and CP-odd Higgs states because of sizeable singlet
admixtures. The heavy MSSM-like Higgs states could have avoided discovery
due to too small cross sections because of missing phase
space for their production or singlet admixtures in their couplings to
the SM particles.\footnote{A further reduction of branching ratios into
  SM particle final states, and hence discovery signatures, can be due
  to possible Higgs-to-Higgs decays.} Figures~\ref{fig:MMM1} show the results of
a scan over a subspace of the NMSSM parameter space:
\begin{eqnarray}
\begin{array}{lll}
1 \le \tan\beta \le 30 \, , \quad & \phantom{-}100 \mbox{ GeV } \le
\mu \le 500
\mbox{ GeV } \, , \quad & 100 \mbox{ GeV } \le M_1 \le 1 \mbox{ TeV }, \\
0.5 \le \lambda \le 0.8 \, , \quad & -500 \mbox{ GeV } \le A_\lambda \le 800
\mbox{ GeV } \, , \quad & 200 \mbox{ GeV } \le M_2 \le 1 \mbox{ TeV },
\\
0 \le \kappa \le 0.8 \, , \quad & -500 \mbox{ GeV } \le A_\kappa \le 200
\mbox{ GeV } \, , \quad & 1.1 \mbox{ TeV } \le M_3 \le 2 \mbox{ TeV },
\end{array}
\end{eqnarray}
and
\begin{eqnarray}
\begin{array}{ll}
- 2 \mbox{ TeV } \le A_U,A_D,A_E \le 2 \mbox{ TeV } &
M_{\tilde{\mu}_R,\tilde{e}_R} = M_{\tilde{L}_{1,2}} =
M_{\tilde{Q}_{1,2}} = M_{\tilde{c}_R,\tilde{u}_R} =
M_{\tilde{s}_R,\tilde{d}_R} = 2.5 \mbox{ TeV } \\
500 \mbox{ GeV } \le M_{\tilde{t}_R} = M_{\tilde{Q}_3} \le 1.5 \mbox{
  TeV } \, ,&
500 \mbox{ GeV } \le M_{\tilde{\tau}_R} = M_{\tilde{L}_3} \le 1 \mbox{
  TeV }\, , \qquad
M_{\tilde{b}_R} = 1 \mbox{ TeV }.
\end{array}
\end{eqnarray}
Here $M_{i}$ ($i=1,2,3$) denote the soft SUSY breaking gaugino masses,
$M_{\tilde{x}_R}$ and $M_{\tilde{X}_j}$ ($j=1,2,3$) the right-handed
and left-handed soft SUSY breaking sfermion masses and $A_{U,D,E}$ the
trilinear soft SUSY breaking couplings of the up- and down-type
quarks and the charged leptons. The thus generated squark and gluino
masses have not been excluded yet by the LHC searches. The scan
leads to scenarios where either the lightest Higgs boson $H_1$ or the second
lightest $H_2$ has a mass of 126~GeV and signal rates that
are compatible within 1$\sigma$ with the rates reported by ATLAS and CMS.
In the following, we present results for the latter case. The scenarios
with a SM-like $H_1$ lead to similar conclusions.
Figure~\ref{fig:MMM1} (left) shows the mass values of the lightest
scalar and pseudoscalar Higgs bosons $A_1$ and $H_1$. The former
ranges between $\sim 62$ and $\sim 480$~GeV, while $M_{H_1} \gesim
18$~GeV with the upper bound given by the $H_2$ mass of 126~GeV. In the
case where $H_1$ and $H_2$ are almost degenerate in mass, the Higgs signal
observed at the LHC is built up by two Higgs bosons. The LEP
exclusion limits are avoided due to $H_1$ and $A_1$ being rather
singlet-like, which leads to suppressed couplings and small signal
rates. The singlet components for $A_1$ and $H_1$ are shown in
Fig.~\ref{fig:MMM1} (right). They are quantified by the corresponding
matrix elements squared, $|P_{A_1 a_s}|^2$ and $|S_{H_1h_s}|^2$, of the
mixing matrices $P$ and $S$,  which rotate the CP-odd and CP-even Higgs
interaction states to the mass eigenstates. As anticipated, $A_1$
and $H_1$ are rather singlet-like. The masses of the heavy
Higgs bosons $A_2$ and $H_3$ are almost degenerate and larger than
$\sim$250~GeV. They are MSSM-like with a singlet admixture
below $\sim 15$ percent. The combination of small phase space and not large
enough couplings implies signal rates which are not in conflict with the
present LHC exclusion limits. Note, that scans over larger ranges of
the NMSSM parameter space lead to similar results. \\

In the future high-luminosity phase of the LHC with 14~TeV c.m. energy
it will be possible to find more than one Higgs boson eventually.
In case three Higgs bosons will be found, precision measurements of
couplings can be exploited to decide whether they are MSSM or NMSSM
Higgs bosons. The NMSSM scalar Higgs boson couplings to gauge bosons
fulfill the following sum rule,
\begin{eqnarray}
\sum_{i=1}^3 g_{H_i VV}^2 = 1 \;,
\end{eqnarray}
and for the couplings to top and bottom quarks we have the sum rule
\begin{eqnarray}
\frac{1}{\sum_{i=1}^3 g^2_{H_i tt}} + \frac{1}{\sum_{i=1}^3 g^2_{H_i
    bb}} = 1
\end{eqnarray}
[in units of SM couplings].
In case only three NMSSM Higgs bosons are discovered, and not all
of them are scalar, the
measurement of their gauge and Yukawa couplings will
show a violation of these sum rules due to the missing couplings of
the non-discovered Higgs bosons. However, in the MSSM, with only
three neutral Higgs bosons, the rules would be fulfilled.
\begin{figure}
\includegraphics[width=0.47\textwidth]{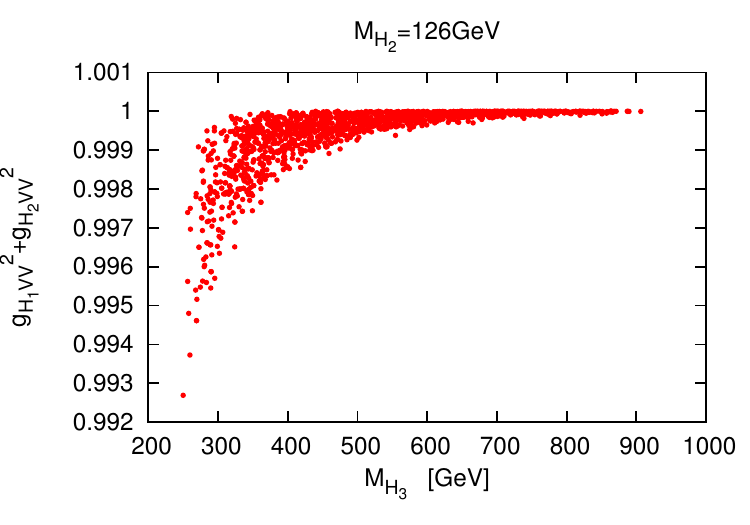}
\includegraphics[width=0.47\textwidth]{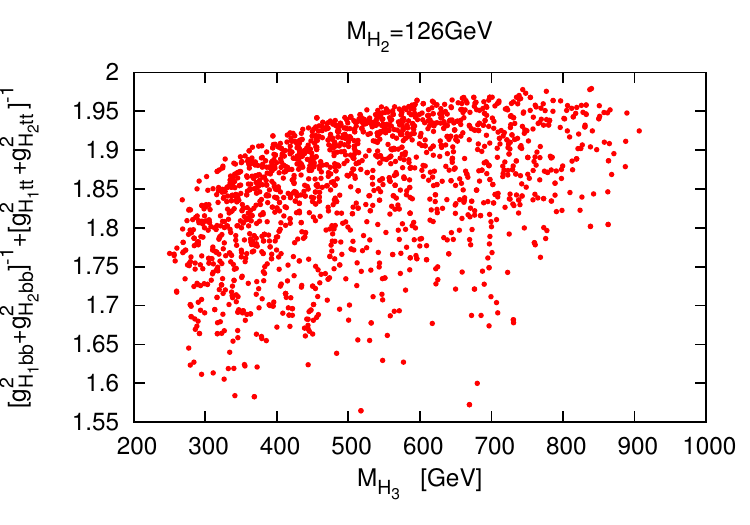}
\vspace{-1.2em}
\caption{Left: Sum of the $H_1$ and $H_2$ (=SM-like) gauge couplings
  squared. Right: Sum of the inverse $H_1$ and $H_2$ Yukawa couplings
  squared. Both are shown as a function of $M_{H_3}$.}
\label{fig:MMM2}
\end{figure}
Figure~\ref{fig:MMM2}
shows the result of a scan where $H_2$ is SM-like and its rates are in
accordance with the measured rates of the LHC experiments. It
is supposed that only the two lightest CP-even Higgs bosons have been
discovered. The left plot shows the violation of the vector coupling
sum rule for $H_1$ and $H_2$, the right plot the violation of the
Yukawa coupling sum rule. The respective sums deviate from 1, in case
of the fermion couplings by up to a factor of
two. Figure~\ref{fig:MMM3} shows the same coupling sums but this time
for the two heavier scalar bosons $H_2$ and $H_3$ and as a function of
$M_{H_1}$, supposing that the
lightest scalar Higgs boson has not been discovered due to too small SM
couplings because of its singlet nature, but the heavier MSSM-like
$H_3$ has been found. The measurement of the Yukawa couplings
would show a violation of the sum rule by up to $\sim 20\%$ in this case, the
gauge couplings by up to $\sim 27$\%. The largest deviations
are observed when the lightest Higgs mass is of ${\cal O}$(100~GeV), close to
the $H_2$ mass, where the two states start to mix strongly. In both scenarios the
determination of the couplings allows to distinguish the NMSSM Higgs
sector from the MSSM provided that the precision in the coupling measurements is
sufficiently high. \\
\begin{figure}
\includegraphics[width=0.47\textwidth]{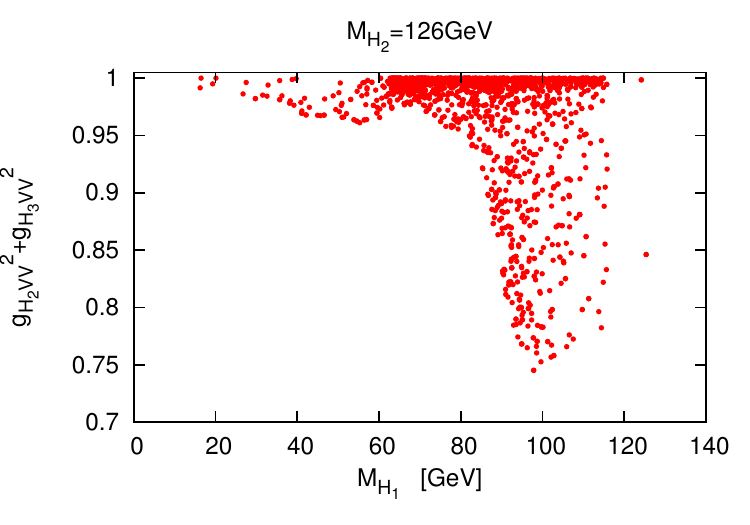}
\hspace*{0.2cm}
\includegraphics[width=0.47\textwidth]{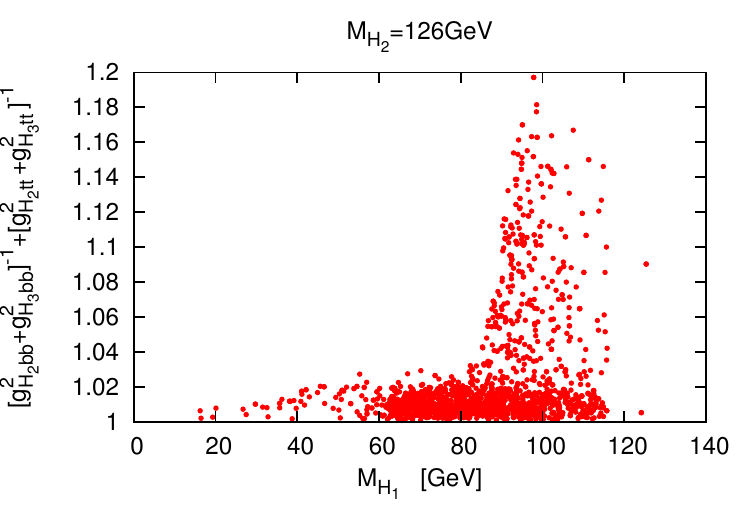}
\vspace{-1.2em}
\caption{Same as Fig.~\ref{fig:MMM2}, but for $H_2$ (=SM-like) and
 $H_3$, and as a function of $M_{H_1}$.}
\label{fig:MMM3}
\end{figure}

In some scenarios, it may be possible to deduce the mass scale of the unobserved
third CP-even scalar from the pattern of violation of the sum rules. For
instance, Fig.~\ref{fig:MMM2} shows that larger deviations for the sum of the
couplings of $H_1$ and $H_2$ to vector bosons are only possible for relatively
small values of $M_{H_3}$, while maximal deviations of the fermion sum rule
point towards larger $H_3$ masses. However, the figure also illustrates that
very different values of $M_{H_3}$ can lead to very similar results for the sum
rules, so that in general it is not possible to reliably infer bounds on the
heavy scale ($M_{H_3}$) from precision measurements of the light observed
scalars. This can be understood from the fact that, contrary to the MSSM, the
NMSSM at tree-level depends on four additional unknown parameters besides
$\tan\beta$ and the charged Higgs boson mass $M_{H^\pm}$\footnote{The charged
Higgs boson mass can be traded for the trilinear soft SUSY breaking coupling
$A_\lambda$.}. These are given by the two NMSSM specific couplings $\lambda$ and
$\kappa$, the vacuum expectation value $v_s$ of the singlet field and the soft
SUSY breaking trilinear coupling $A_\kappa$. The additional parameters influence
the values of the mixing angles and thus the Higgs boson couplings to the light
scalar particles. Therefore there is no unique correlation between the coupling
values and the scale of new physics, as given {\sl e.g.} by the charged Higgs
boson mass.\\

In the NMSSM the Higgs signal observed at the LHC could also be built up by two
Higgs bosons, which are nearly degenerate in mass
\cite{nmssmscans,Gunion:2012gc}, while such scenarios are difficult
to achieve in the MSSM. In case no further Higgs bosons are discovered, the
observation of a Higgs signal built up by two resonances clearly
allows to distinguish the NMSSM from the SM case. The superposition of two
(or more) nearly degenerate Higgs bosons near 126~GeV can be
tested experimentally by analyzing double ratios of signal rates
\cite{Gunion:2012he}. The deviation from unity in this case could be
tested at the 14~TeV run of the LHC with high luminosity.

\section{Loop effects \label{sec:loopeffects}}

Due to the numerical loop coefficient $1/(16\pi^2)$ and potentially
small couplings, loop effects are less promising for probing energy scales with
BSM physics, in particular because experimental analyses partly rely on tails of
distributions. In general the mass range probed would be covered by
direct searches for new particles at the LHC [and later LC]. Nevertheless,
certain extensions of the Standard Model, such as supersymmetric models, generate
parametrically enhanced loop corrections. In addition, somewhat exotic
configurations, such as leptophilic particles equipped only with new
$U(1)$ charges, may be suppressed in production modes at the LHC, and thus
precision Higgs analyses may open a window to new BSM degrees of
freedom.

\subsection{Simple examples}
\label{sec:examples}

Obviously, we cannot cover loop effects in all different Higgs production and decay
modes, so that
we focus on three representative cases: first, we discuss vertex corrections to
fermionic Higgs decays, mediated by heavy gauge and scalar bosons. They are similar, in spirit,
to the well studied corrections to $Z \to b\bar{b}$ decays in many models. Next, we
introduce corrections to the loop-induced Higgs--gluon and Higgs--photon interactions,
which benefit from the fact that the leading Standard Model amplitudes are already
loop-suppressed. Finally, we show some distinctive effects of vector-like leptons.

\subsubsection*{Heavy virtual bosons}
\label{sec:loop_bosons}

\begin{figure}[b]
\centering
\includegraphics[width=0.18\textwidth, bb=224 283 441 499]{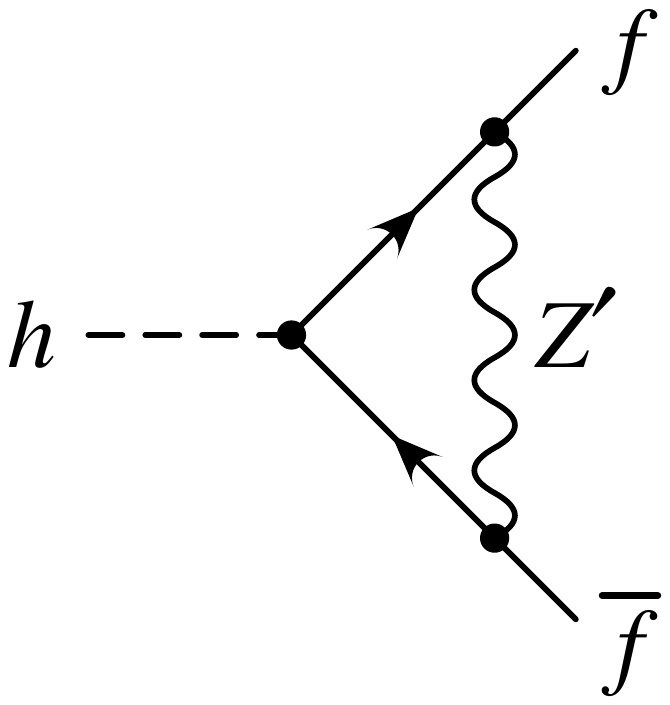} \hspace{30mm}
\includegraphics[width=0.18\textwidth, bb=224 283 441 499]{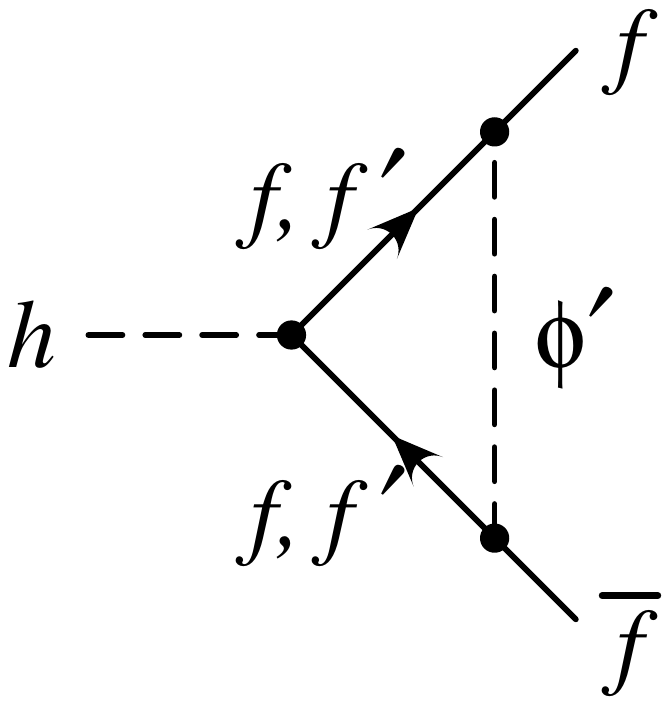}
\caption{Feynman diagrams for Higgs--fermion
coupling corrections from vector boson or scalar exchange.}
\label{fig:feyab}
\end{figure}

Higgs vertex corrections due to $Z'$ vector boson and $\sigma'$
scalar exchanges, as shown in
Fig.~\ref{fig:feyab},
have a generic structure and a general
magnitude of loop effects. In terms of the shift of the Higgs coupling
to light fermions $g \to (1+\Delta)\, g $, one finds for a vector exchange:
\begin{equation}
   \Delta_{Z'}	    =	    - \frac{g_L g_R}{4 \pi^2}\, m^2_h\, C_0(m^2_h,0,0,0,0,m^2_{Z'})
		    \,\simeq\,	 - \frac{g_L g_R}{4 \pi^2}\,
 \frac{m^2_h}{m^2_{Z'}} \left[\log \frac{m^2_h}{m^2_{Z'}} - 1\right] \,,
 \label{vcorr}
\end{equation}
where the last step corresponds to the limit $m_{Z'} \gg m_h$.
The couplings $g_{L/R}$ denote the left-chiral and right-chiral
gauge couplings of the $Z'$ boson to the light SM
fermions.

To preserve gauge invariance, the scalar $\sigma'$ must belong to an
$SU(2)$ doublet $\phi'$, but it is assumed not to be a Higgs boson
\textsl{i.e.} not carry a vacuum expectation value. The leading
correction to the $hb\bar{b}$ coupling reads, in the limit $m_{\phi'}
\gg m_t \gg m_b$:
\begin{equation}
   \Delta_{\phi'}   \simeq  \frac{y_U y_D y_t}{64 \pi^2 y_b}\,
    \frac{m^2_h}{m^2_{\phi'}}
    \left[ 4\tau^3 \arctan \frac{1}{\tau} - 2\tau^2\log\frac{m^2_{\phi'}}{m^2_t}
    -2\tau^2 -1 \right]
      \quad \text{with} \quad \tau = (4 m_t^2 / m^2_h -1)^{1/2}  \label{scorr} \,,
\end{equation}
where $y_{D,U}$ are the Yukawa couplings for the $\bar{q}_L \phi' b_R$
and $\bar{q}_L \phi'^c t_R$ interactions, respectively.  The result
can be adapted straightforwardly for other fermion final states.
We see that both corrections, Eq.\eqref{vcorr} and Eq.\eqref{scorr} decrease with
the square of the new physics mass scale, as expected qualitatively from the
operator expansion and in accordance with the decoupling theorem~\cite{appcar}.
Setting the couplings $g_L g_R = 1$ (or $y_Uy_Dy_t/y_b = 1$), the
magnitude of the corrections $\Delta$ is of order $10^{-2}$ or less
for $m_{Z'} \gesim 300$~GeV ($m_{\phi'} \gesim 300$~GeV), so
significant deviations are expected only in exotic scenarios with
large couplings.

\subsubsection*{Loop-induced decays}
\label{sec:loopi}

The effective Higgs--gluon and Higgs--photon couplings are excellent
probes for new physics because already in the Standard Model they are
loop-induced. The only difference between the top and $W$ loops in the
Standard Model and new physics contributions is the actual mass suppression
described in Eq.\eqref{eq:coup_rescale}. While in $e^+ e^-$ collisions
Higgs decays to gluons can be extracted from the backgrounds, at the LHC the sensitive
observable to the Higgs--gluon coupling is the production cross section,
as described in detail in Section~\ref{sec:eff}. In the following we will
focus on Higgs decays, but all results can be directly translated into
the LHC production rate.
Of course, the $h \to \gamma Z$
process is also a loop-induced interaction, but after including the
$Z\to\ell^+\ell^-$ branching fraction the expected event rate is very
low, so it is challenging to precisely measure this channel at
colliders. Therefore the $\gamma Z$ mode is in general less suitable for
constraining generic new physics effects and thus will not be discussed
here.
\medskip

The SM results for loop-induced Higgs decays~\cite{gg,gg1} can be written as
\begin{alignat}{7}
\Gamma[h \to gg] &= \frac{\alpha_\text{s}^2m_h^3}{128\pi^3}|{\cal A}_{gg}|^2\,,
\qquad
& {\cal A}_{gg} &= (\sqrt{2}G_F)^{1/2}\sum_q A_{1/2}(\tau_q)\,,
\notag \\
\Gamma[h \to\gamma\gamma] &= \frac{\alpha^2m_h^3}{1024\pi^3}
 |{\cal A}_{\gamma\gamma}|^2\,,
& {\cal A}_{\gamma\gamma} &= (4\sqrt{2}G_F)^{1/2} \left[
 \sum_f N_c^f Q_f^2 A_{1/2}(\tau_f) + A_1(\tau_W)\right] \,, \label{eq:delaa}
\end{alignat}
where $G_F$ is the Fermi constant, $N_c^f =1$,3 for leptons ($f=
\ell$) and quarks ($f=q$),
respectively, and $\tau_i = 4m_i^2/m_h^2$. The loop functions are defined as
\begin{alignat}{7}
A_{1/2}(\tau) &= 2\tau [1+(1-\tau)f(\tau)]\,, \notag \\
A_1(\tau) &= -[2+3\tau+3\tau(2-\tau)f(\tau)]\,, \label{eq:Afunc}
\end{alignat}
where
\begin{alignat}{7}
f(\tau) &= \left\{ \begin{array}{ll}
\arcsin^2 \sqrt{1/\tau}\,, & \tau \geq 1, \\[.5ex]
-\dfrac{1}{4}\Bigl [ \log \dfrac{1+\sqrt{1-\tau}}{1-\sqrt{1-\tau}} -i\pi
\biggr ]^2, & \tau < 1.
\end{array}\right.
\end{alignat}
A new heavy particle $X$ that couples to the Higgs boson with strength
$g_{hXX}^{}$ leads to an additional contribution to these rates given
by~\cite{Gillioz:2013pba,newgg,newggnlo}
\begin{alignat}{7}
\Delta{\cal A}_{gg} &= \frac{g_{hXX}^{}}{m_X^2}
 T_X\delta_R \left\{
 \begin{array}{ll} A_0(\tau_X) & \text{for scalar }X, \\
 2m_XA_{1/2}(\tau_X) & \text{for fermion }X, \\
 A_1(\tau_X) & \text{for vector }X,
 \end{array}\right. \label{hggmod} \\
\Delta{\cal A}_{\gamma\gamma} &=
 \frac{g_{hXX}^{}}{m_X^2} N^X_{c}Q_X^2 \left\{
 \begin{array}{ll} A_0(\tau_X) & \text{for scalar }X, \\
 2m_XA_{1/2}(\tau_X) & \text{for fermion }X, \\
 A_1(\tau_X) & \text{for vector }X,
 \end{array}\right. \label{haamod}
\end{alignat}
where $T_X = 0,\frac{1}{2},3$ and $N^X_c=1,3,8$ if $X$ is a QCD singlet,
triplet or octet, respectively. Furthermore, $\delta_R = 1/2$ for a
self-conjugate field and $\delta_R = 1$ otherwise, and
\begin{alignat}{7}
A_0(\tau) &= -\tau [1-\tau f(\tau)].
\end{alignat}
Higher-order QCD corrections to the $hgg$ interaction are known to be
large, but they have been calculated to complete NNLO and partial NNNLO
order~\cite{hggsm} in the Standard Model within the heavy top quark limit,
and to NLO for the generic new physics contributions in Eq.\eqref{hggmod}
\cite{newgg2} in the case of large loop particle masses. For
estimating the sensitivity to heavy new physics effects, it is however
sufficient to consider the tree-level formulae listed above, since
bottom quark effects and mismatches of the relative QCD corrections
between the different contributions range at the 10--20\% level.

For large values of $m_X$, the loop functions simplify to
$A_0=\frac{1}{3}$, $A_{1/2}=\frac{4}{3}$, and $A_1 = -7$. In this limit,
the contributions in Eq.\eqref{eq:delaa} can be mapped onto the $D=6$
operators in Table~\ref{tab:D6op} according to
\begin{alignat}{7}
f_{GG} &= -\frac{g_{hXX}^{}}{96\pi^2v}
 T_X\delta_R \left\{
 \begin{array}{ll} 1 & \text{for scalar }X, \\
 8m_X & \text{for fermion }X, \\
 -21 & \text{for vector }X,
 \end{array}\right.  \label{hggcoeff} \\
\tfrac{1}{2}(f_{BB}+f_{WW}-f_{BW}) &=
 -\frac{g_{hXX}^{}}{96\pi^2v} N^X_{c}Q_X^2 \left\{
 \begin{array}{ll} 1 & \text{for scalar }X, \\
 8m_X & \text{for fermion }X, \\
 -21 & \text{for vector }X .
 \end{array}\right. \label{haacoeff}
\end{alignat}
Note that electroweak gauge symmetry demands that the coupling between
$h$ and a fermion $X$ emerges from a dimension-5 operator of the form
$\frac{1}{\Lambda}\phi^\dagger\phi \bar{X}X$, so that $g_{hXX}^{}$ in this case
should be of order $\ope(v/\Lambda)$. Assuming a common new physics scale
$\Lambda \sim m_X$, one can thus see that the
expressions in eqs.~(\ref{hggcoeff},\ref{haacoeff}) are independent of $m_X$ for
all spin assignments of $X$.

\subsubsection*{Vector-like leptons}
\label{sec:vecl}

An enlarged spectrum that feeds into modifications of the $h\to
\gamma\gamma$ branching ratio is typically accompanied by a
modification of the $h\to Z\gamma$ branching. While the effects on
$h\to Z\gamma$ can be larger ({\sl{e.g.}} in composite scenarios
\cite{Azatov:2013ura}), a measurement at the LHC can be challenging. A
different avenue to formulate constraints on such a situation is via
precision measurements at a future lepton collider. An enlarged
spectrum modifies the $e^+e^-\to h Z$ production cross section through
higher-order electroweak corrections, which can be significantly
larger in the BSM theory. Such a cross section modification, which can
be studied in a model-independent fashion at a future 250~GeV lepton
collider, might indeed be resolvable given the high precision
measurements that can be performed with such a
machine~\cite{lsb,Englert:2013tya,Craig:2013xia}.
The uncertainties that
arise at hadron colliders render such a measurement more difficult at
the LHC, but nevertheless an important first step is
possible. Depending on the precision of the coupling extraction that
can be obtained at the LHC and the performance of, {\sl{e.g.}},
boosted analyses of $pp\to hZ$ production~\cite{signal_b},
deviations in the high $p_T$ region due to resolved loop contributions
can be used to formulate bounds on non-SM contributions as discussed
in~\cite{resolveloop}.

\begin{figure}[t]
\centering
\includegraphics[scale=0.7]{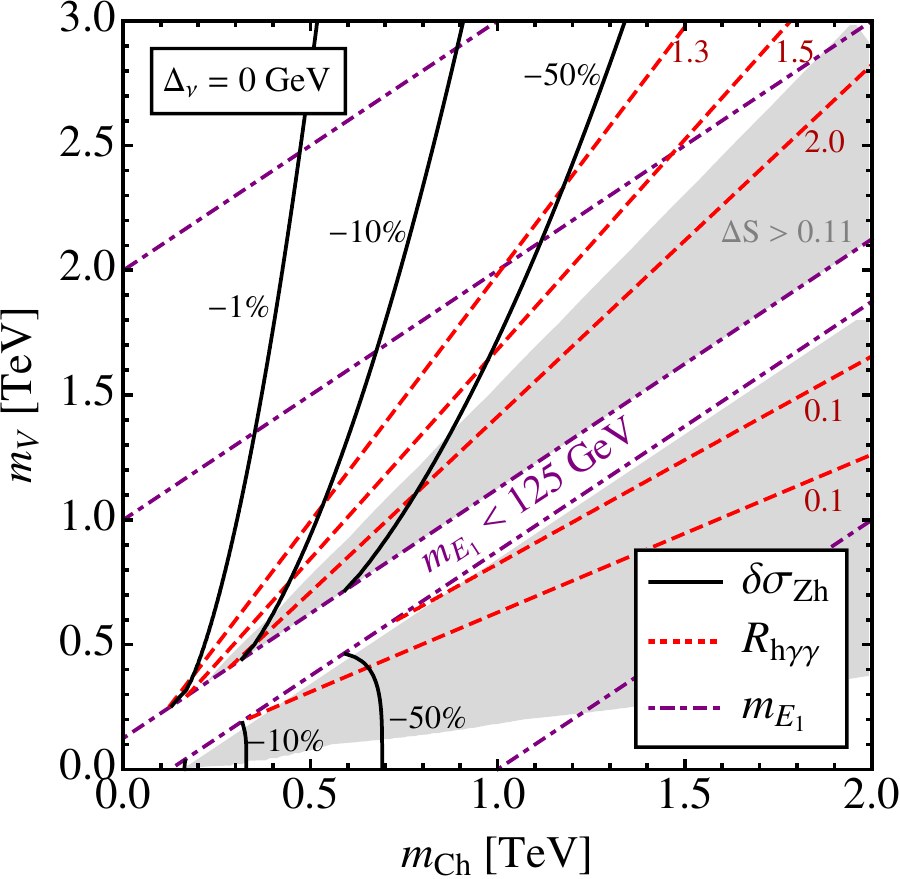}
\hskip 0.8cm
\includegraphics[scale=0.7]{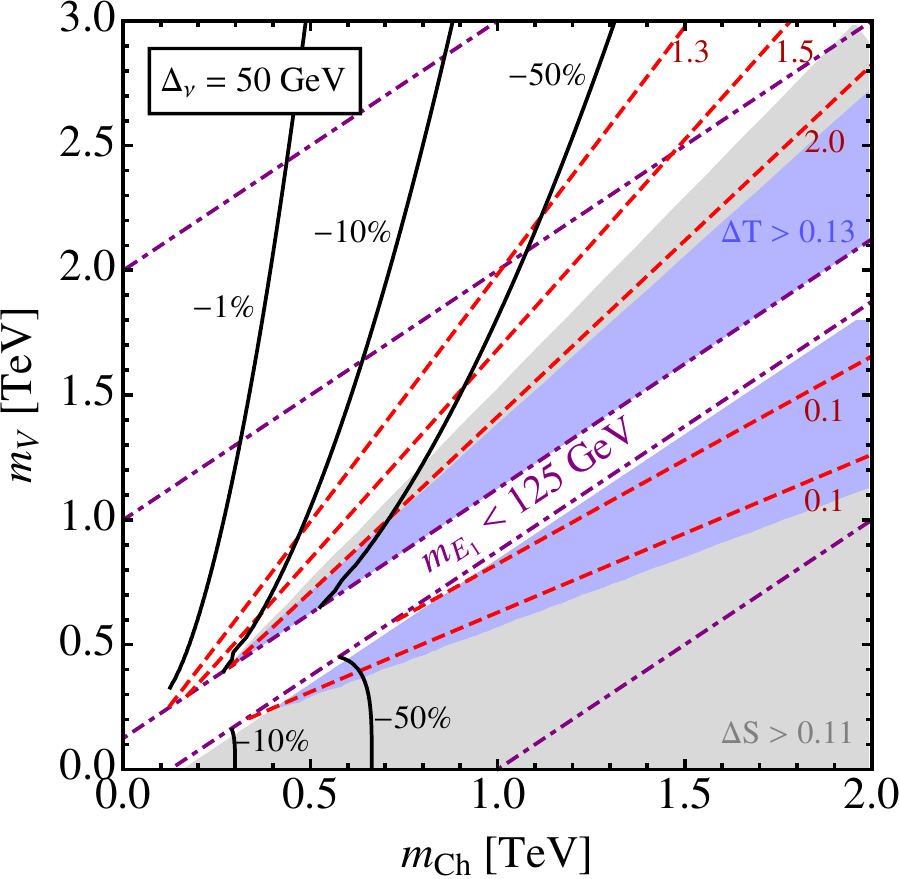}
\caption{Correlation of the diphoton branching modification (red
  dashed) and the modification of the associated Higgs production
  cross section (solid black) that arises as part of the NLO
  electroweak correction. Parameter regions that are excluded by
  oblique correction constraints on $\Delta T, \Delta S$ are
  shaded. The parameter region that results in a lightest charged
  lepton eigenvalue $E_1$ lighter than 125 GeV is excluded. The left plot
  corresponds to $\Delta_\nu=0$, while in the right plot $\Delta_\nu=50$~GeV.
  Figure taken from Ref.~\cite{Englert:2013tya}. \label{fig:deltasig}}
\end{figure}

In the following we will focus on these effects in the clean environment
of a linear collider experiment, where they can be resolved best.
Electroweak corrections are most straightforward in a
well-defined framework. As a concrete model to review the precision
measurement avenue we consider a simple scenario of vector-like
leptons as discussed in~\cite{Joglekar:2012vc}
\begin{alignat}{5}
  \label{eq:vector}
  -{\cal L}\,\, &\supset m_\ell \bar{\ell}'_L {\ell }''_R + m_e
  \bar{e}''_L {e}'_R
  + m_\nu \bar{\nu}_L'' {\nu}_R' + \text{h.c.}	 \\
  \label{eq:higgs}
  & +Y_c' (\bar{\ell}'_L h ) {e}_R' + Y_n' (\bar{\ell}'_L i\sigma^2
  h^\dagger) {\nu}_R' + Y_c'' (\bar{\ell}''_R h ) {e}_L''
  + Y_n'' (\bar{\ell}''_R i\sigma^2 h^\dagger)	{\nu}_L'' + \text{h.c.}\notag
\end{alignat}
where $\ell'_L, \ell''_R=({\bm{2}},-1/2)$, $e''_L,e'_R=({\bm{1}},-1)$,
and $\nu''_L,\nu'_R=({\bm{1}},0)$ under $SU(2)_L\times U(1)_Y$. To
reduce the number of free parameters we choose common values for the
vector-like lepton masses
\begin{equation}
  m_\ell=m_e=m_\nu = m_V
\end{equation}
and common ``chiral'' masses from the couplings to the Higgs vev
\begin{alignat}{5}
  Y_c' v/ \sqrt{2} & = Y_c'' v/ \sqrt{2} = m_{Ch} \nonumber \\
  Y_n' v/ \sqrt{2} & = Y_n'' v/ \sqrt{2} = m_{Ch}+\Delta_\nu ~~,
  \label{eq:VL4hcoup}
\end{alignat}
in the following. Depending on these mass parameters
\cite{lsb,Englert:2013tya,Joglekar:2012vc}, the $h\to \gamma\gamma$
branching fraction can be enhanced via Eq.\eqref{haamod},
Fig.~\ref{fig:deltasig}, while direct collider constraints and
precision measurements are currently not sensitive to such a spectrum
(see Ref.~\cite{Joglekar:2012vc} for a discussion). Obviously the
potentially large enhancement of the $h\to \gamma\gamma$ branching
ratio is a way to constrain the chiral component of this model at the
LHC once the $h\to \gamma \gamma$ becomes SM-like. However, the
percent-level precision measurement of the associated production cross
section at a linear collider can supersede the LHC measurement as
indicated in Fig.~\ref{fig:deltasig}.

Contextualizing Fig.~\ref{fig:deltasig} with the aim of our review, we
can identify a region $m_{Ch}\sim v$ in \eqref{eq:VL4hcoup} and make a
prediction on the vectorial mass terms (the new mass scale in this
concrete model) given a precision percent-level measurement at a
future lepton collider. Figure~\ref{fig:deltasig} shows that the new
physics scale is in this case at around 1.5 TeV.

\subsection{$\tan\beta$-enhanced non-decoupling effects}
\label{sec:stb}

While generic effects of weakly interacting extended Higgs sectors can
be expected to decouple with the mass scale of the heavy new states,
this does not have to be true for all models. In general extended Higgs
models there exists a universal source of such effects: the mass of the
heavy new state might receive contributions both directly from the Higgs
potential Eq.\eqref{eq:2hdm_pot} and from a combination of a vev and a
self interaction, $m^2_{H^0} \sim m_{12}^2 + \lambda_3 v_1^2/2$.
The second term appears as a result of spontaneous symmetry breaking
and avoids the Appelquist--Carazzone decoupling
theorem~\cite{appcar} when $v_1$ and thus $m_{H^0}$ become large. It can
lead to substantial corrections to the effective quartic Higgs
coupling, but also to the other Higgs couplings. More
generally, large contributions of the kind $\lambda_i v^2$ can
induce unexpectedly large effects, because they are only constrained by
the high-scale behavior of the extended Higgs model.

In general Two-Higgs-Doublet models the most dangerous source of large
quantum corrections is related to the form of the top Yukawa coupling to
the heavy states: dependent on the model setup it scales with
$m_t/\tan \beta$, which means that for $\tan \beta < 1$ it rapidly
approaches a Landau pole. Experimentally, such effects are strongly constrained
for example by $B_d-\overline{B}_d$ mixing.
\medskip

\begin{figure}[tb]
\centering
(a)\hspace{-1em}
\includegraphics[height=0.17\textwidth, bb=72 434 594 568]{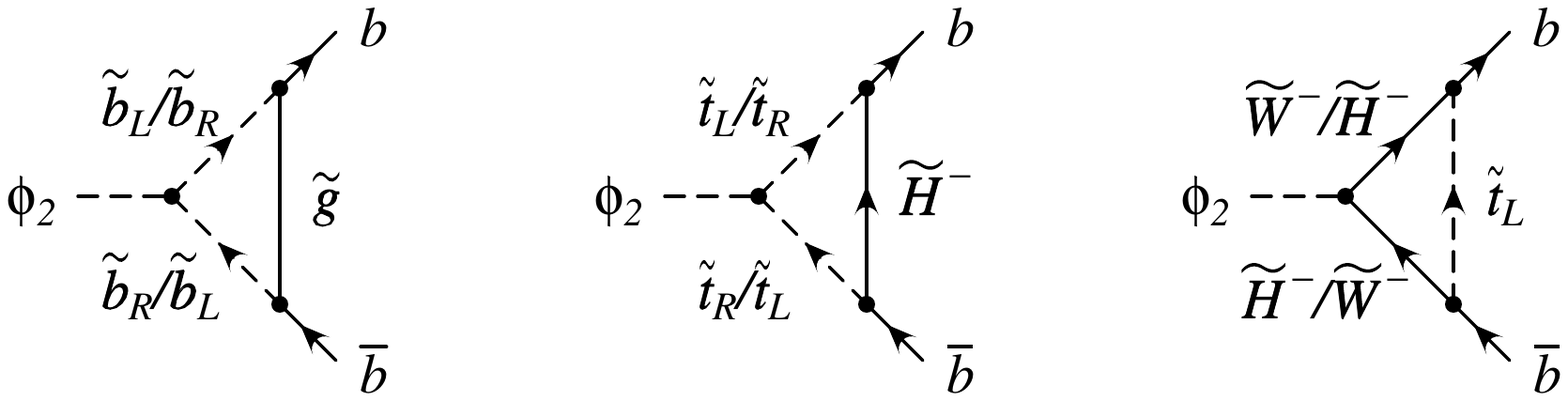}
\hspace{15mm}
(b)\hspace{-1em}
\includegraphics[height=0.17\textwidth, bb=72 434 234 568]{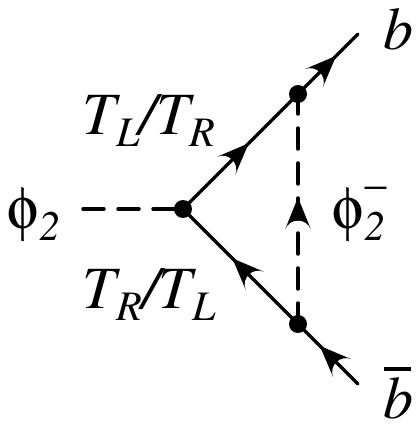}
\caption{Feynman diagrams leading to $\tan\beta$-enhanced corrections to
the bottom-quark Yukawa coupling in (a) the MSSM and (b) the model
\eqref{darkyuk} with heavy vector-like fermions. Here $\phi_2$
is the second Higgs doublet, and
$\tilde{b}$, $\tilde{t}$, $\tilde{g}$, $\tilde{W}^-$, and $\tilde{H}^-$
are the sbottom, stop, gluino, wino and higgsino fields, respectively.
}
\label{fig:feydb}
\end{figure}

In the type-II 2HDM of the MSSM, a subset of the radiative corrections
to down-type Yukawa couplings, $y_d$, are enhanced by $\tan\beta$ and do
not decouple in the limit $\mu \sim M_\text{SUSY} \gg v$~\cite{deltamb}.
These corrections emerge from loop contributions to the down-type
quark-Higgs coupling involving the insertion of the up-type
Higgs field $\phi_2$ instead of the down-type Higgs field $\phi_1$,
see Fig.~\ref{fig:feydb}~(a). This calculation can be reduced to the
corresponding self-energies with the insertion of the vev
$v_2$ and using its relation to the full Higgs field $\phi_2$ by means of a
low-energy theorem \cite{gg}. The leading terms are
proportional to the QCD
coupling $\alpha_\text{s}\tan\beta$ or the top Yukawa coupling
$y_t\tan\beta$ and thus can be $\ope(1)$. For the $h^0b\bar{b}$ coupling
they can be written, in the limit $\mu \sim M_\text{SUSY} \gg v$, as
\begin{alignat}{7}
&\frac{\Gamma_\text{SUSY}[h^0\to bb]}{\Gamma_\text{Born}[h^0\to bb]} =
\left[\frac{1}{1+\Delta_b} \left(1-\frac{\Delta_b}{\tan\alpha\tan\beta}\right)\right]^2 \,,
\notag \\
&\Delta_b =
-\frac{2\alpha_\text{s}\tan\beta}{3\pi}\,\frac{\mu}{m_{\tilde{g}}} \,
 I\left( \frac{m_{\tilde{b}_L}^2}{m_{\tilde{g}}^2},
	 \frac{m_{\tilde{b}_R}^2}{m_{\tilde{g}}^2} \right)
-\frac{y_t^2\tan\beta}{16\pi^2}\,\frac{A_t}{\mu} \,
 I\left( \frac{m_{\tilde{t}_L}^2}{\mu^2},
	 \frac{m_{\tilde{t}_R}^2}{\mu^2} \right)
+\frac{3g^2\tan\beta}{32\pi^2}\,\frac{M_2}{\mu} \,
 I\left( \frac{m_{\tilde{t}_L}^2}{\mu^2},
	 \frac{M_2^2}{\mu^2} \right)  \notag \\
&I(x,y) = \frac{x\log x}{(1-x)(x-y)} + \frac{y\log y}{(1-y)(y-x)}\,.
\label{eq:dmb}
\end{alignat}
Here $\mu$ and $M_2$ are the higgsino and wino mass parameters,
respectively; $m_{\tilde{b}_{L,R}}$, $m_{\tilde{t}_{L,R}}$ and
$m_{\tilde{g}}$ denote the masses of the sbottoms, stops and gluino,
respectively; and $y_t$ is the top Yukawa coupling. Note that for the
leading contribution in $\tan\beta$, mixing between the L- and R-sfermions
and among the charginos can be neglected.

Besides the terms listed in Eq.\eqref{eq:dmb} there are additional
contributions proportional to $g'^2$ which, however, are suppressed by
the small hypercharge of the bottom quark and thus negligible. The
leading two-loop corrections to these effective Yukawa couplings have
been determined in Ref.~\cite{deltamb2}. They are of moderate size if
the scales of the QCD coupling $\alpha_s$ and the top Yukawa coupling
$y_t$ are chosen as the average of the correspondingly contributing
SUSY masses.\medskip

While the occurrence of non-decoupling $\tan\beta$-enhanced corrections is
natural in the MSSM or other supersymmetric models~\cite{nmssmcalc},
they can more generally
appear in a large class of models that include the 2HDM type-II and a set of new
particles at the mass scale $\Lambda$. The crucial condition is that large
mixing between the up- and down-type Yukawa sectors is generated through loops
of $\ope(\Lambda)$ fields. [In the MSSM this mixing is achieved by the $\mu$
term.]

For example, we may consider a type-II 2HDM supplemented by a dark (hidden) sector that
is charged under a $\mathbb{Z}_2$ symmetry. The dark sector contains vector-like
fermionic partners of the top quark, denoted $Q_L = (T_L,\,B_L)$ and $T_R$, as
well as a scalar partner of the SM-like Higgs doublet $\phi_1$,
denoted $\phi_2$, with Yukawa couplings of the form
\begin{equation}
{\cal L}_\text{dark,yuk} = -x_d y_b\,\overline{Q}_L\phi_2 b_R -
 x_u y_t\,\overline{q}_{3L}\phi_2^c T_R - y_t\,\overline{Q}_L\phi_1^c T_R
 + \text{h.c.}\,,
\label{darkyuk}
\end{equation}
where $q_{3L} = (t_L,\,b_L)$ is the left-handed SM quark doublet, and $x_u$ and
$x_d$ are arbitrary ${\cal O}(1)$ parameters.
It is assumed that the fields $Q_L$, $T_R$ and $\phi_2$ all have masses $m_{Q_L} \sim m_{T_R} \sim
m_{\phi_2} \sim\Lambda$.
The dark sector may contain additional fields, but they are unnecessary for
our purposes. Since $\phi_2$ couples to both up- and down-type quarks in
eq.~\eqref{darkyuk}, it generates the required mixing between these two
sectors.
As a result, the loop diagram involving $T_L$, $T_R$, and the charged
component of $\phi_2$, see Fig.~\ref{fig:feydb}~(b), leads to the
$\tan\beta$-enhanced correction
\begin{alignat}{7}
\Delta_b = \frac{y_t^2 x_u x_d \tan\beta}{4\pi^2}
	 + \ope \left( \frac{v^2}{\Lambda^2} \right) \; .
\end{alignat}
Generically, therefore, a large deviation with respect to the Standard Model, of
the Higgs couplings to down-type fermions may be interpreted as a hint
for the presence of a second Higgs doublet and a hidden sector.

\section{Summary and Conclusions}
\label{sec:sum}

The discovery of the Higgs boson not only concludes the search
  for the fundamental degrees of freedom predicted by the Standard Model, but also
  provides us with the unique opportunity to gain information about
  anticipated new physics through a detailed study of its
  properties.
We have evaluated the sensitivity of
Higgs precision measurements to various new physics scenarios. The main findings
are summarized in Table~\ref{tab:reach}, based on the following input:

\begin{table}[tbp]
\begin{tabular}{|l||c|c||c|c|}
\hline
Scenario/framework		   &  LHC  & HL-LHC  & LC    & HL-LC \\
\hline\hline
Higgs portal			   &  0.23 & 0.28    & 0.44  & 0.56  \\
2HDM type-II ($\tan\beta\approx1 $)&  0.52 & 0.58    & 1.15  & 1.6   \\
2HDM type-II ($\tan\beta\approx10$)&  0.33 & 0.36    & 0.7   & 1.0   \\
\hline
$D=6$ effective operators: &&&& \\
$\quad hVV$			   &  0.78 & 0.87    & 2.6   & 3.3   \\
$\quad hff$			   &  0.45 & 0.50    & 1.0   & 1.4   \\
$\quad hgg$ contact		   &  0.55 & 1.1     & 1.3   & 1.8   \\
$\quad h\gamma\gamma$ contact	   &  0.15 & 0.18    & 0.24  & 0.36  \\
\hline
Strong interactions		   &  0.9  & 1.1--2.0&2.8--5.1&3.4--5.6 \\
\hline
$hgg$ loop effects: &&&& \\
$\quad$scalar triplet		  &  0.16 & 0.31    & 0.37  & 0.52  \\
$\quad$scalar octet		 &  0.39 & 0.75    & 0.92  & 1.3   \\
$\quad$vector octet		 &  1.8  & 3.5	   & 4.2   & 5.8   \\
$h\gamma\gamma$ loop effects: &&&& \\
$\quad$scalar triplet		  &  0.15 & 0.18    & 0.24  & 0.36  \\
$\quad$scalar octet		 &  0.25 & 0.29    & 0.39  & 0.60  \\
$\quad$vector octet		 &  1.1  & 1.3	   & 1.8   & 2.7   \\
\hline
Vector-like leptons		  &  ---  &  ---    & 1.2   & 1.5   \\
\hline
\end{tabular}
\caption{Effective sensitivity to new physics scales (in TeV) at the level of one standard
deviation from measurements of the Higgs couplings at the LHC and
a future LC (with different assumptions about the luminosity, see
Table~\ref{tab:cplgs}) in a variety of BSM models.}
\label{tab:reach}
\end{table}

\begin{itemize}
\item \underline{\it Higgs portal:} Eq.\eqref{eq:portalcpl} combined with
Table~\ref{tab:cplgs};
\item \underline{\it Two-Higgs-Doublet Model:} Eq.\eqref{eq:thdm_fact} combined
with Table~\ref{tab:cplgs};
\item \underline{\it effective dimension-6 operators:} Results from
Table~\ref{tab:Lambdas};
\item \underline{\it Strongly interacting Higgs field:} Results from
Table~\ref{tab:xitab};
\item \underline{\it Heavy virtual bosons:} No competitive limits;
\item \underline{\it Loop-induced decays:} Eqs.\eqref{hggmod},\eqref{haamod}
with $\delta_R=1$, $Q_X=1$ and Table~\ref{tab:cplgs}
\item \underline{\it Vector-like leptons:} Fig.~\ref{fig:deltasig} for
$m_{Ch}\approx v$.
\end{itemize}
In deriving the limits quoted in the table, we have generically assumed that the
couplings between the Higgs boson and the new physics sector are $\ope(1)$ for
gauge-invariant dimensionless couplings or $\ope(v)$ for couplings with mass
dimension one.

Evidently, Higgs precision data can be sensitive to multi-TeV
scales, beyond the reach of direct LHC searches as well as electroweak
precision tests. This is true in particular if the new physics \textit{(i)}
modifies the Higgs couplings to $W$ and $Z$ bosons, \textit{(ii)} is related to
strong dynamics, or \textit{(iii)} involves new heavy vector bosons that
generate loop corrections to the Higgs--gluon or Higgs--photon interactions. On the
other hand, in minimal weakly coupled scenarios such as the Higgs portal or a
scalar triplet loop in Higgs--gluon or Higgs--photon couplings, the Higgs coupling measurements
only probe sub-TeV scales, and thus complement direct LHC searches.

Observation of significant deviations in future Higgs coupling precision
measurements, combined with data from direct searches for new particles at the
LHC and precision electroweak and flavor constraints, will allow one to
narrow down the possibilities for new physics that can explain the
Higgs coupling shifts. The pattern of deviations in the different couplings
carries additional information. For instance, extended scalar sectors may lead to
large deviations in the Higgs--fermion couplings, as illustrated in
eq.~\eqref{eq:thdm_fact} and in Figs.~\ref{fig:MMM2}. On the other hand,
sizeable shifts in the Higgs--gauge boson couplings can be generated, for
instance, by strong interactions.

Most notably, in supersymmetric theories and models with similar particle
content, there can be large radiative corrections to the coupling between the Higgs
boson and down-type quarks that do not decouple for large values of the mass
scale of the particles in the loop. Such a scenario offers sensitivity to
extremely high scales from Higgs precision physics, although it may be
fine-tuned from a model building perspective.\medskip

If multiple Higgs scalars are directly observed at the LHC, the
measurement of their couplings will again yield valuable information about
potential extra unobserved states. In particular, the discovery of three neutral
Higgs bosons would suggest a Two-Higgs-Doublet model as the underlying theory,
as for example in the MSSM, but they may also be part of a more complex scalar
sector, involving for instance an additional singlet, as in the NMSSM. These two
cases may be distinguished by analyzing the couplings of the three observed
states to gauge bosons and fermions, which would violate certain sum rules if
the additional particles are not completely decoupled. However, the mass scale
of the extra scalars (beyond the two Higgs doublets) can in general not be
deduced from this information, since any extension of the
Two-Higgs-Doublet model introduces several new parameters, which may lead to partial
cancellations in the corrections to the couplings.

In either of these two scenarios, with only one observed Higgs boson or multiple
Higgs bosons at the LHC, precision measurements of its couplings provide a
unique window into possible new physics sectors, which are not excluded by
current data. In order to fully exploit this avenue, percent-level precision for
the determination of Higgs production and decay rates will be essential, a goal
that can be achieved by a future high-energy $e^+e^-$ collider, such as
the ILC.

\subsection*{Acknowledgements}

The authors are deeply indebted to P.~M.~Zerwas, who has initiated this review
and has provided extensive advise during its preparation. We also
thank C.~Grojean and A.~Pomarol for helpful discussions.
C.E.\ is supported in parts by the IPPP Associateship programme.
A.F.\ acknowledges support by the National Science Foundation under
grant no.\ PHY-1212635. M.M., M.R.\ and K.W.\ are supported by the DFG
SFB/TR9 ``Computational Particle Physics''. M.R. additionally
acknowledges support by the BMBF under Grant No. 05H09VKG
(``Verbundprojekt HEP-Theorie'').


\end{document}